%% file: main.tex
\documentclass[pageno]{jpaper}
\usepackage{mathptmx} 

\usepackage[normalem]{ulem}

\usepackage{fancyhdr}
\usepackage[normalem]{ulem}
\usepackage[hyphens]{url}
\usepackage[sort,nocompress]{cite}
\usepackage{flushend}

\usepackage{xspace}
\usepackage{makecell}
\usepackage{subfigmat}

\usepackage[bookmarks=true,breaklinks=true,colorlinks,
pdfpagelabels=false,linkcolor=black,citecolor=blue,urlcolor=black]{hyperref}
\usepackage[capitalize,nameinlink]{cleveref}
\usepackage{caption}
\DeclareCaptionFont{9pt}{\fontsize{9pt}{10pt}\selectfont}
\newcommand{\system}{\textsc{Tvarak}\xspace}
\newif\ifcommenton
\commentontrue
\ifcommenton
\newcommand{\rajat}[1]{{\color{magenta}{\textsf [Rajat: #1]}}}
\newcommand{\nathan}[1]{{\color{red}{\textsf [Nathan: #1]}}}
\newcommand{\greg}[1]{{\color{blue}{\textsf [Greg: #1]}}}
\else
\newcommand{\rajat}[1]{}
\newcommand{\nathan}[1]{}
\newcommand{\greg}[1]{}
\fi 

\title{\system: Software-managed hardware offload for DAX NVM storage redundancy}
\author{Rajat Kateja, Nathan Beckmann, Gregory R. Ganger\\rkateja@cmu.edu, beckmann@cmu.edu, ganger@ece.cmu.edu\\Carnegie Mellon University\vspace{-3mm}}

\begin{document}
\date{}
\maketitle
\thispagestyle{empty}


\input{abstract}
\input{intro}

\input{background}

\input{design}

\input{eval}

\input{conclusion}

\pagebreak
\bibliographystyle{plain}
\bibliography{references}

\end{document}

%% file: abstract.tex
\begin{abstract}
\system efficiently implements system-level redundancy
for direct-access (DAX) NVM storage.
Production storage systems complement device-level ECC (which covers
media errors) with 
system-checksums and cross-device parity.
This system-level redundancy enables detection of
and recovery from data corruption due to device
firmware bugs 
(e.g., reading data from the wrong physical location).
Direct access to 
NVM penalizes 
software-only implementations of system-level 
redundancy, forcing a choice between lack of data
protection
or 
significant 
performance penalties.
Offloading the update and verification of system-level redundancy
to \system, a hardware controller co-located with the last-level cache, 
enables efficient protection of data 
from such bugs in memory controller and NVM DIMM firmware.
Simulation-based evaluation with seven data-intensive applications
shows \system{}'s performance and energy efficiency.
For example, \system reduces Redis set-only performance by only 3\%,
compared to 50\% reduction for a state-of-the-art software-only approach.
\end{abstract}

%% file: intro.tex
\section{Introduction}
\label{sec:intro}

Non-volatile memory (NVM) 
storage 
improves the performance of stateful applications 
by offering 
DRAM-like performance with disk-like 
durability~\cite{bpfs-sosp, pmfs-eurosys, 
nova-fast, nstore-sigmod, wbl-vldb}. 
Applications that seek to leverage raw NVM performance  
eschew conventional file system and block interfaces 
in favor of direct access (DAX) to NVM. With DAX, an application 
maps NVM data into its address space and uses load and store 
instructions to access it, eliminating system software 
overheads from the data path~\cite{linux-dax, mnemosyne-asplos, 
nvheaps-asplos, pmfs-eurosys, aerie-eurosys}.

Production storage systems protect data from
various failures.
In addition to fail-stop failures like 
machine or device crashes, storage 
systems also need to 
protect data from silent corruption 
due to firmware bugs. 
Storage device firmware is prone to 
bugs because of its complexity, and these 
bugs can cause data corruption. 
Such corruption-inducing firmware bugs fall into two broad 
categories: lost write bugs and misdirected 
read or write bugs~\cite{ironfs-sosp, data-corruption-tos, 
storage-subsystem-failures-tos, 
long-term-digital-storage-eurosys, 
data-integrity-techniques-applications-storagess}. 
Lost write bugs cause the firmware to acknowledge a write 
without ever updating the data on the device media.
Misdirected read or write 
bugs cause the firmware to read or write the 
data from the wrong location on the device media. 
Firmware-bug-induced corruption will go unnoticed
even in the presence of device-level ECC, 
because that ECC is read/written as an atom with its data during
each media access performed by the firmware.
%

Protection against firmware-bug-induced corruption commonly relies on
system-checksums for detection 
and cross-device parity for recovery. 
\emph{System-checksums} are data checksums that the 
storage system computes and verifies at a layer 
"above" the device firmware (e.g., the file system), 
using separate I/O requests
than for the corresponding data~\cite{ironfs-sosp, 
data-corruption-tos, nova-fortis-sosp}.
Using separate I/O requests for the data and the block containing
its system-checksum (together with system-checksums for other data)
reduces the likelihood of an undetected firmware-bug-induced corruption.
This is because 
a bug is unlikely to affect both in a consistent manner. 
Thus, the storage system can detect a firmware-bug-induced 
corruption because of a mismatch between the two.
It can then trigger recovery using the cross-device 
parity~\cite{raid-sigmod, 
designing-for-disasters-fast, ibm-raim, raim-isca}.
In this paper, we use the term \emph{redundancy} to refer to 
the combination of system-checksums and cross-device parity. 

Production NVM-based storage systems will need such redundancy
mechanisms for the same reasons as conventional storage.
NVM device firmware involves increasingly complex functionality,
akin to that of other storage devices, 
making it susceptible to both lost write and misdirected read/write bugs.
However, most existing NVM storage system designs provide insufficient
protection.
Although fault-tolerant NVM file systems ~\cite{nova-fortis-sosp,
plexistore-nvmw} efficiently cover
data accessed through the file system interfaces, they do not 
cover DAX-mapped data. The Pangolin~\cite{pangolin-atc} 
library is an exception, implementing system-checksums 
and parity for applications that use its transactional library 
interface. However, software-only approaches for DAX NVM 
redundancy incur significant performance overhead
(e.g., 50\% slowdown for
a Redis set-only workload, even with Pangolin's streamlined design).



This paper proposes
\system\footnote{\system means accelerator in Hindi.},
a software-managed hardware offload
that efficiently maintains redundancy for DAX NVM data.
\system co-resides with the last level cache (LLC) controllers
and coordinates with the file system to provide 
DAX data coverage without application involvement. 
The file system informs \system when it DAX-maps a file.
\system verifies each DAX NVM cache-line read and 
updates the redundancy for each DAX NVM cache-line 
write-back. 

\system's design relies on two key elements to achieve
efficient redundancy verification and updates. 
First, \system reconciles the mismatch between DAX 
granularities (typically 64-byte cache lines) and typical 4KB
system-checksum block sizes
by introducing cache-line granular system-checksums 
(only) while data is DAX-mapped. 
\system accesses these cache-line granular system-checksums,
which are themselves packed into cache-line-sized units,
via separate NVM accesses. Maintaining these checksums 
only for DAX-mapped data limits the resulting space overhead. 
%
Second, \system uses caching to reduce the number of extra NVM accesses
for redundancy information.
Applications' data access locality 
leads to reuse of system-checksum and 
parity cache-lines; \system leverages this reuse with a 
small dedicated on-controller cache and configurable 
LLC partitions for redundancy information. 

Simulation-based evaluation with seven applications, each 
with multiple workloads, demonstrates \system{}'s promise of
efficient DAX NVM storage redundancy. 
For Redis, \system incurs only a 3\% 
slowdown for maintaining redundancy with a 
set-only workload, in contrast to 50\% slowdown with 
Pangolin's efficient software approach,
without compromising on coverage or checks.
For other applications and workloads, the results consistently
show that \system efficiently updates and verifies 
system-checksums and parity,
especially in comparison to
software-only alternatives.
The efficiency benefits are seen in both application runtimes 
and energy. 

This paper makes three primary contributions.
First, it motivates the need for architectural support for DAX NVM storage
redundancy, highlighting the limitations of software-only approaches.
Second, it proposes \system{}, a low-overhead, software-managed 
hardware offload for DAX NVM storage redundancy. It describes the challenges for efficient
hardware DAX NVM redundancy and how \system overcomes these challenges with
straightforward, effective design.
Third, it reports on extensive evaluation of \system{}'s runtime, energy,
and memory access overheads for seven applications, each under multiple
workloads, showing its efficiency especially in comparison to
software-only alternatives.

%% file: background.tex
\section{Redundancy Mechanisms and NVM Storage}
\label{sec:background}

This section provides background and discusses related work.
First, it describes conventional storage 
redundancy mechanisms for firmware bug resilience. 
Second, it discusses the need for these mechanisms in
NVM storage systems, the direct-access (DAX) interface to 
NVM storage, and the challenges in maintaining 
the required redundancy with DAX.
Third, it discusses related work and where \system fits.

\subsection{Redundancy for Firmware Bug Resilience}
\label{sec:redundancy-mechanisms}
Production storage systems employ a variety of redundancy 
mechanisms to address a variety of
faults~\cite{zfs-end-to-end-integrity-fast, gfs-sosp, 
z2fs-msst, wafl-integrity-fast, pvfs-checksums-pdsw, 
nova-fortis-sosp, raid-sigmod,   
snapmirror-fast, designing-for-disasters-fast, 
seneca-atc}. 
In this work, we focus on redundancy mechanisms
used to detect and recover from firmware-bug-induced 
data corruption (specifically, per-page system-checksums and cross-device parity).

\begin{figure}[t]
	 \begin{subfigmatrix}{1}
		 \subfigure[The problem: device responds to read of block
                 that experienced the lost write with incorrect (old) data.
                 ]{\label{fig:lost-write}
			\includegraphics[width=\columnwidth, keepaspectratio]{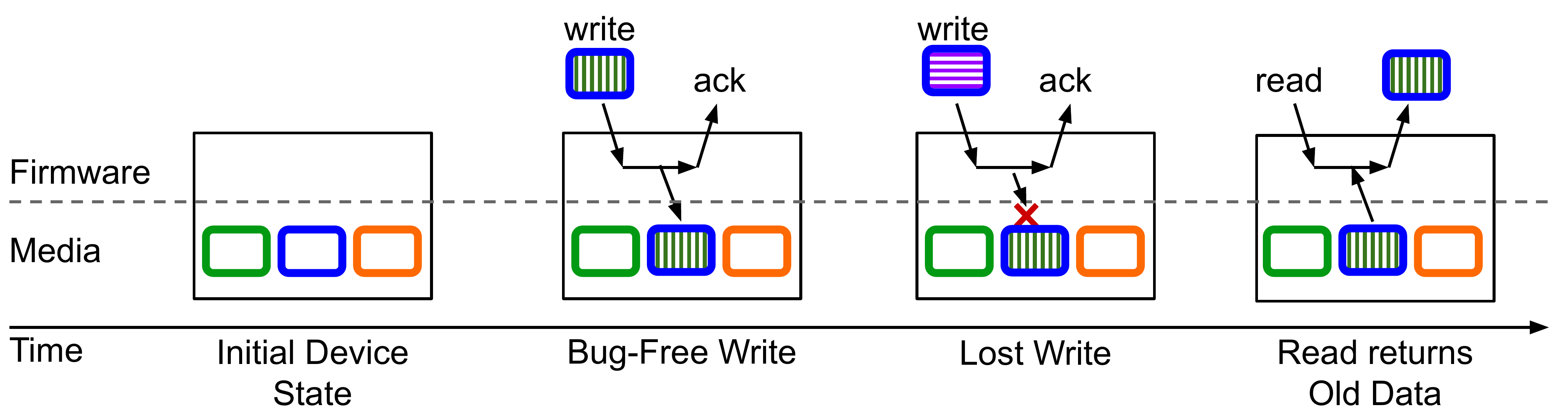}}
		\subfigure[The fix: having the higher-level system update
                and verify system-checksums when writing or reading data,
		in separate requests to the device, enables detection of a lost write 
		because of mismatch between the data and the system-checksum.]{\label{fig:lost-write-detection}
			\includegraphics[width=\columnwidth, keepaspectratio]{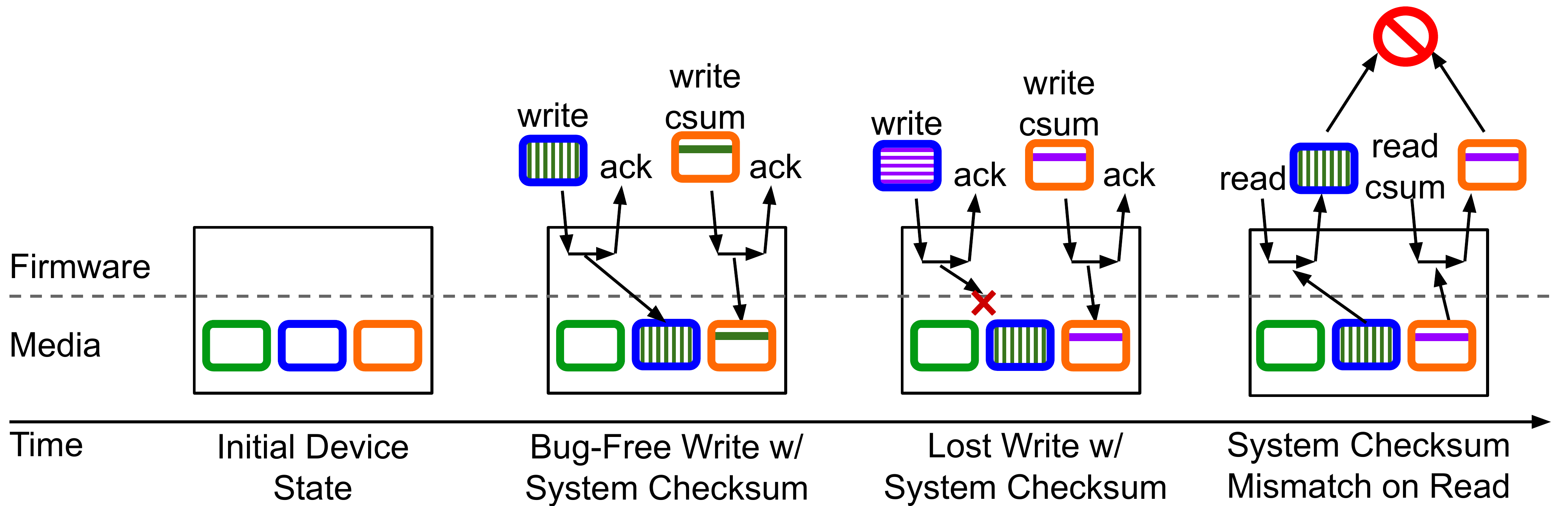}}
	 \end{subfigmatrix}
	\caption{Lost write bug example.  Both sub-figures show a time-line
		for a storage device with three media locations. 
		The device is shown in an initial state, and then upon 
		completion of higher-level system's write or read to data 
		(first, a successful write, then a "lost write", 
		then a read) mapped to the same media location.
		(a) shows how 
		the higher-level system can consume incorrect (old) data 
		if it trusts the device to never lose
		an acknowledged write.
		(b) shows how the higher-level system can detect a lost write with
		system-checksums.
	 }
\label{fig:lost-write-example}
\end{figure}

\textbf{Firmware-bug-induced data corruption}: 
Large-scale studies of deployed storage systems 
show that device firmware bugs sometimes
lead to data loss or corruption~\cite{ironfs-sosp, data-corruption-tos, 
storage-subsystem-failures-tos, 
long-term-digital-storage-eurosys, 
data-integrity-techniques-applications-storagess}.
Device firmware, like any software, is prone to bugs because 
of its complex responsibilities (e.g., address 
translation, dynamic re-mapping, wear leveling, block caching,
request scheduling) 
that have increased both in number 
and complexity over time.
Research has even proposed 
embedding the entire file system functionality~\cite{devfs-fast}
and application-specific functionalities~\cite{active-storage-data-mining-vldb, 
active-virtual-objects-hotstorage, active-disks-data-processing-ieee, 
active-flash-ics, active-ssd-search-engine}
in device firmware.
Increasing firmware complexity increases the propensity 
for bugs, some of which can trigger data loss or corruption.

Corruption-inducing 
firmware-bugs can be categorized into two broad categories: 
lost write bugs and misdirected read/write bugs.
A lost write bug causes the firmware to acknowledge a write 
without ever updating the media with the write 
request's content. An example scenario that can 
lead to a lost write is if a write-back firmware cache "forgets"
that a cached block is dirty.
\cref{fig:lost-write} illustrates a lost write bug.
It first shows (second stage in the time-line) a correct bug-free
write to the block stored in the blue media location.
It then shows a second write to the same block, but this one
suffers from a lost write bug---the firmware acknowledges the write
but never updates the blue media location. 
The subsequent read of the 
blue block returns the old data to the application.

\begin{figure}[t]
	 \begin{subfigmatrix}{1}
		 \subfigure[The problem: device responds to read with the
 incorrectly updated data from the blue location.  Notice that the green location 
 also has incorrect (old) after the misdirected write.
			]{\label{fig:misdirected-write}

			\includegraphics[width=\columnwidth, keepaspectratio]{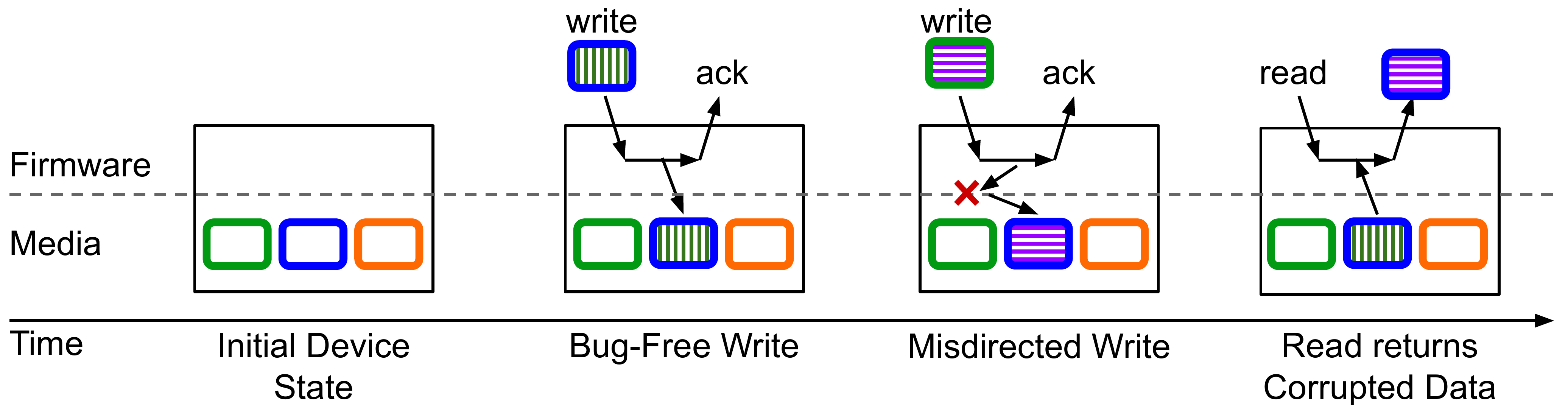}}
		\subfigure[The fix: having the higher-level system update
                and verify system-checksums when writing or reading data,
                in separate requests to the device, enables
		 enables detection of a misdirected write 
		because of mismatch between the data and the system-checksum.]
		{\label{fig:misdirected-write-detection}
			\includegraphics[width=\columnwidth, keepaspectratio]{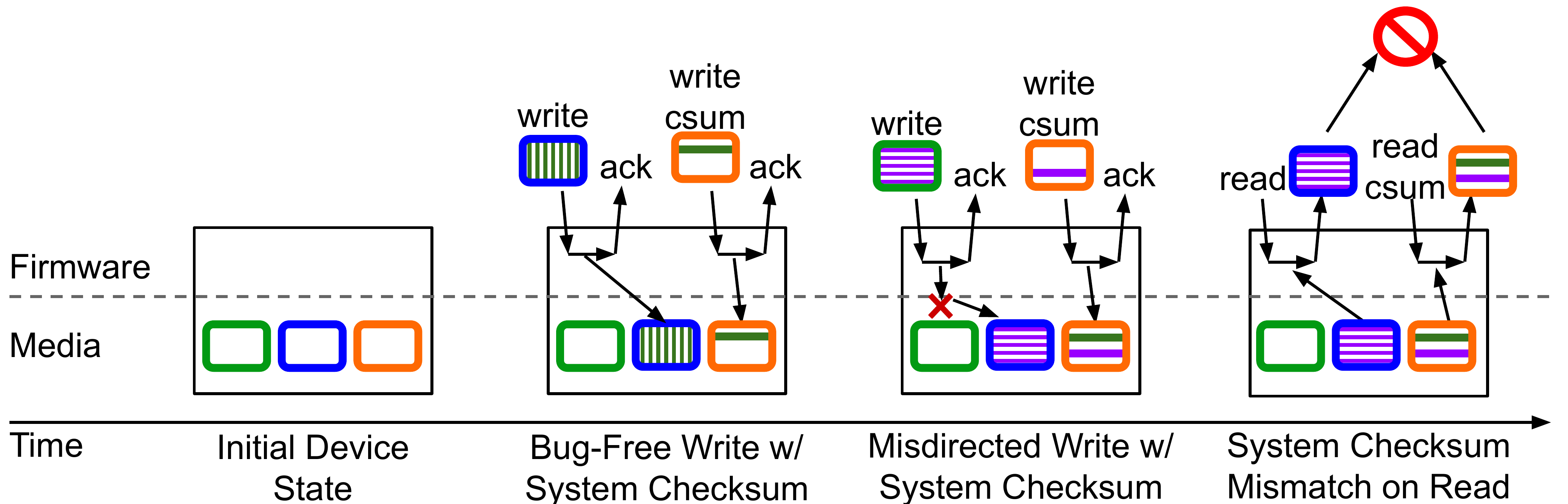}}
	 \end{subfigmatrix}
	 \caption{Misdirected write bug example.  Similar construction
         to \cref{fig:lost-write-example}, but with the second operation
         being a write intended for the green location that is misdirected
         by the firmware to the blue location.
         }
\label{fig:misdirected-write-example}
\end{figure}

A misdirected write or misdirected read bug causes the 
firmware to store data at or 
read data from an incorrect media location, 
respectively.
\cref{fig:misdirected-write} illustrates a misdirected write bug.
As before, the first write to the block stored in the blue location
is performed correctly by the firmware.
For this example, the second write request shown is for the block stored
in the green location.
But, it encounters a misdirected write bug wherein the 
data is incorrectly written to the blue media location. 
Notice that a misdirected write bug not only fails to update the
intended block, but also corrupts (incorrectly replaces) the data of the block
it incorrectly updates.
In the example, the subsequent read to the the block mapped
to the blue location returns this corrupted data. 

Although almost all storage devices maintain
error-correcting codes (ECCs) 
to detect corruption due to random bit 
flips~\cite{chipkill-whitepaper, lot-ecc-isca, chipkill-arcc-hpca, 
nvm-ecc-micro}, 
these ECCs cannot detect firmware-bug-induced 
corruption~\cite{ironfs-sosp, data-corruption-tos}. 
Device-level ECCs are stored together with the data and 
computed and verified inline by the same firmware
during the actual media update/access.
So, in the case of a lost write, the firmware loses the ECC update 
along with the corresponding data update, because the 
data and ECC are written together on the media as one operation.
Similarly, misdirected writes modify the 
ECC to match the incorrectly updated data and 
misdirected reads retrieve the ECC corresponding to the 
incorrectly read data. 


\textbf{System-checksums for detection}:
Production storage systems maintain 
per-page \emph{system-checksums} to detect 
firmware-bug-induced data corruption. System-checksums are 
updated and verified at a layer above the 
firmware, such as the file system, 
stored in checksum blocks (each containing checksums for many blocks)
separate from the data, 
and read and written using I/O requests separate
from the corresponding data I/O 
requests~\cite{zfs-end-to-end-integrity-fast, gfs-sosp, 
z2fs-msst, wafl-integrity-fast, pvfs-checksums-pdsw, 
nova-fortis-sosp}.
Separating the storage and accesses for 
data from corresponding system-checksums
enables detection of 
firmware-bug-induced corruption,
because such bugs are unlikely to affect both.
The probability of a bug affecting 
both in a consistent fashion (e.g., losing both 
or misdirecting both to another corresponding 
data and system-checksum pair) is even lower.

\cref{fig:lost-write-detection} demonstrates how 
system-checksums enable detection of lost writes. 
Although the second write to the blue block is lost, the 
write to the checksum block (stored in the orange location) is not.
Thus, upon the data read in the example, which is paired with a 
corresponding system-checksum read and verification, 
the lost write is detected. 

\cref{fig:misdirected-write-detection} illustrates 
how system-checksums enable detection of misdirected writes.
A misdirected write firmware bug is extremely unlikely to 
affect both the data write to the green block 
and the corresponding system-checksum write to the orange block
in a consistent manner.
To do so, the firmware would have to 
incorrectly write the system-checksum to a location 
(block \emph{and} the offset within the block)
that stores the checksum for the exact block to which it 
misdirected the data write. 
In the illustration, the read of the blue block data, followed 
by its system-checksum read, results in a verification failure. 
Similarly, system-checksums also trigger a verification 
failure in case of a misdirected read bug, because a bug 
is unlikely to affect the 
both the data its system-checksum read. 

\textbf{Cross-device parity for recovery}:
To recover from a detected page corruption, storage systems store
parity pages~\cite{raid-sigmod, 
	snapmirror-fast, designing-for-disasters-fast, 
seneca-atc, raim-isca, ibm-raim}.
Although parity across arbitrarily selected pages suffice 
for recovery from firmware-bug-induced corruption, storage systems 
often implement cross-device parity that enable 
recovery from device failures as well.

\subsection{NVM Storage Redundancy and Direct Access (DAX)}
\label{sec:nvm}
Non-volatile memory (NVM) refers to a class of 
memory technologies 
that have DRAM-like access latency and granularity but 
are also durable like disks~\cite{3dxp, memristor,   
pcm-isca, pcm-wear-leveling-isca, intel-nvm-announcement}.
NVM devices have orders of magnitude lower latency 
and higher bandwidth than conventional storage devices, 
thereby improving stateful applications' 
performance~\cite{bpfs-sosp, 
pmfs-eurosys, nova-fast, scmfs-sc, 
nstore-sigmod, wbl-vldb, foedus-sigmod}.

\textbf{Need for firmware-bug resilience in NVM storage}:
NVM storage systems will be  
prone to firmware-bug-induced data corruption 
and require corresponding redundancy mechanisms,
like conventional storage systems.
NVM firmware is susceptible to 
corruption-inducing bugs, because it 
is non-trivial and its complexity 
can only be expected to increase over time.
NVM firmware already provides for address translation, bad block 
management, wear leveling, request scheduling, and 
other conventional firmware 
responsibilities~\cite{secure-wear-leveling-hpca, 
pcm-wear-leveling-isca, startgap-micro, security-refresh-isca}. 
Looking forward, its complexity will only increase as more 
NVM-specific functionality is embedded into the 
firmware (e.g., reducing NVM writes and wear~\cite{deuce-asplos, 
nvm-dedup-micro, janus-isca, flip-n-write-micro, 
nvm-redundancy-bit-writes-date}) 
and as the push towards near-data 
computation~\cite{3dmemory-pointer-chasing-iccd, pim-instructions-isca, nda-hpca, 
pim-graph-processing-isca, top-pim-hppdc, 
devfs-fast, active-storage-data-mining-vldb, 
active-virtual-objects-hotstorage, active-disks-data-processing-ieee, 
active-flash-ics, active-ssd-search-engine} 
continues.

\begin{table*}[t]
    \centering
{\small
\input{tables/related-work.tex}
}
    \vspace*{10pt} 
    \caption{Trade-offs among \system{} and previous DAX NVM storage redundancy designs.}
    \label{table:related-work}
\end{table*}

\textbf{Direct access NVM storage redundancy challenges}:
Direct-access (DAX) interface to NVM storage 
exposes raw NVM performance to applications~\cite{linux-dax, 
pmfs-eurosys, aerie-eurosys, nova-fortis-sosp, 
pmemcached-hotstorage, 
redis-pmem, mojim-asplos,pmse, pangolin-atc, 
hotpot-socc, lazy-redundancy-pdltr}. 
DAX-enabled file systems map NVM-resident 
files directly into application address spaces;
such direct mapping is possible because 
of NVM's DRAM-like access characteristics.
DAX enables applications to access persistent data with 
load and store instructions, eliminating
system software overheads from the data path. 

These characteristics, however,
pose challenges for maintaining 
firmware-bug resiliency mechanisms~\cite{lazy-redundancy-pdltr}.
First, the lack of interposed system software in the data path
makes it difficult to efficiently identify data reads and writes 
that should trigger a system-checksum verification and 
system-checksum/parity updates, respectively.
Second, updating and verifying system-checksums for DAX data 
incurs high overhead because of the mismatch 
between DAX's fine-grained accesses and the
typically large blocks (e.g., 4KB pages) 
over which checksums are computed for space 
efficiency.


\subsection{Related Work on DAX NVM Storage Redundancy}
\label{sec:motivation}

Existing proposals for maintaining 
system-checksums and parity in NVM storage systems 
compromise on 
performance, coverage, and/or programming flexibility for 
DAX-mapped data.
\cref{table:related-work} summarizes these trade-offs. 
Two recent fault-tolerant NVM file systems, Nova-Fortis~\cite{nova-fortis-sosp}
and Plexistore~\cite{plexistore-nvmw}, update and check redundancy
during explicit FS calls but do not update or verify redundancy 
while data is DAX mapped.
Interposing library-based solutions,
such as  Mojim~\cite{mojim-asplos}, HotPot~\cite{hotpot-socc}, 
and Pangolin~\cite{pangolin-atc}, can protect DAX-mapped data
if applications use the given library's transactional interface for all
data accesses and updates. 
But, software-based redundancy updates on every data update incur
large performance overhead.
Mojim~\cite{mojim-asplos} and 
HotPot~\cite{hotpot-socc} would incur very high overhead because 
of DAX's fine-grained writes~\footnote{The original 
	Mojim and HotPot designs do not include checksums, 
	only replication, but their designs extend naturally
	to include per-page checksums.}.
Pangolin~\cite{pangolin-atc} reduces such overhead 
by eschewing per-page checksums 
in favor of per-object checksums, accepting
higher space overhead instead, but still incurs performance
overheads do to redundancy updates/verifications in software.
Anon~\cite{lazy-redundancy-pdltr} reduces the performance overhead,
potentially arbitrarily,
by delaying 
and batching the per-page checksum updates. 
In doing so, however, Anon reduces the coverage
guarantees by introducing windows of vulnerability wherein 
data can get corrupted silently. 

Most existing redundant NVM storage system designs do 
not verify DAX application data reads with the corresponding checksum.
As shown in the fourth column of Table~\ref{table:related-work},
some do no verification while data is DAX-mapped, while others'
designs would only accommodate verifying checksums as part of
background scrubbing.
Pangolin does verify 
the checksums when it reads an object into a DRAM buffer, 
providing significantly tighter verifications.

The remainder of this paper describes and evaluates \system,
a software-managed hardware controller that provides 
in-line redundancy maintenance
at low overheads and without programming restrictions based on required
use of a given library's interface. 
\system updates the redundancy for every write to 
the NVM device and verifies system-checksums for every 
read from the NVM device.

%% file: tables/related-work.tex
\setlength{\tabcolsep}{3pt}

\begin{tabular}{r||l|l|l|l}
	\makecell[r]{\textbf{NVM Storage Redundancy Design}} & 
	\makecell[l]{\textbf{Checksum}\\\textbf{Granularity}} & 
	\makecell[l]{\textbf{Checksum/Parity Update}\\\textbf{for DAX data}} & 
	\makecell[l]{\textbf{Checksum Verification}\\\textbf{for DAX data}} & 
	\makecell[l]{\textbf{Performance}\\\textbf{Overhead}} \\ 
\hline
	\makecell[r]{Nova-Fortis~\cite{nova-fortis-sosp}, Plexistore~\cite{plexistore-nvmw}} & 
	\makecell[l]{Page} &
	\makecell[l]{No updates} & 
	\makecell[l]{No verification} & 
	\makecell[l]{None} \\
\hline
	\makecell[r]{Mojim~\cite{mojim-asplos}, HotPot~\cite{hotpot-socc}} & 
	\makecell[l]{Page$^{2}$} &
	\makecell[l]{On application data flush} & 
	\makecell[l]{Background scrubbing} & 
	\makecell[l]{Very High} \\
 \hline
	\makecell[r]{Pangolin~\cite{pangolin-atc}} & 
	\makecell[l]{Object} &
	\makecell[l]{On application data flush} & 
	\makecell[l]{On NVM to DRAM copy} & 
	\makecell[l]{High} \\
 \hline
	\makecell[r]{Anon~\cite{lazy-redundancy-pdltr}} & 
	\makecell[l]{Page} &
	\makecell[l]{Periodically} & 
	\makecell[l]{Background scrubbing} & 
	\makecell[l]{Configurable} \\
 \hline 
	\makecell[r]{\system} & 
	\makecell[l]{Page} & 
	\makecell[l]{On LLC to NVM write} & 
	\makecell[l]{On NVM to LLC read} & 
	\makecell[l]{Low}\\
\end{tabular}

%% file: design.tex
\section{\system Design}
\label{sec:design}

\newcommand{\Tcsums}{DAX-CL-checksums\xspace}
\newcommand{\tcsums}{DAX-CL-checksums\xspace}
\newcommand{\tcsum}{DAX-CL-checksum\xspace}

\system is a hardware controller that is co-located 
with the last-level cache (LLC) bank controllers. 
It coordinates with the file system to 
protect DAX-mapped data from firmware-bug-induced 
corruptions. 
We first outline the goals of \system. 
We then start by describing a naive redundancy 
controller design, and improve its design to 
reduce its overheads, leading to \system's design. 
We end with \system's architecture, 
area overheads, and walk through examples. 

\subsection{\system's Goals and Non-Goals}
\label{sec:goals}
\begin{figure}[t]
	\centering
	\vspace*{10mm}
	{\includegraphics[width=0.7\columnwidth]{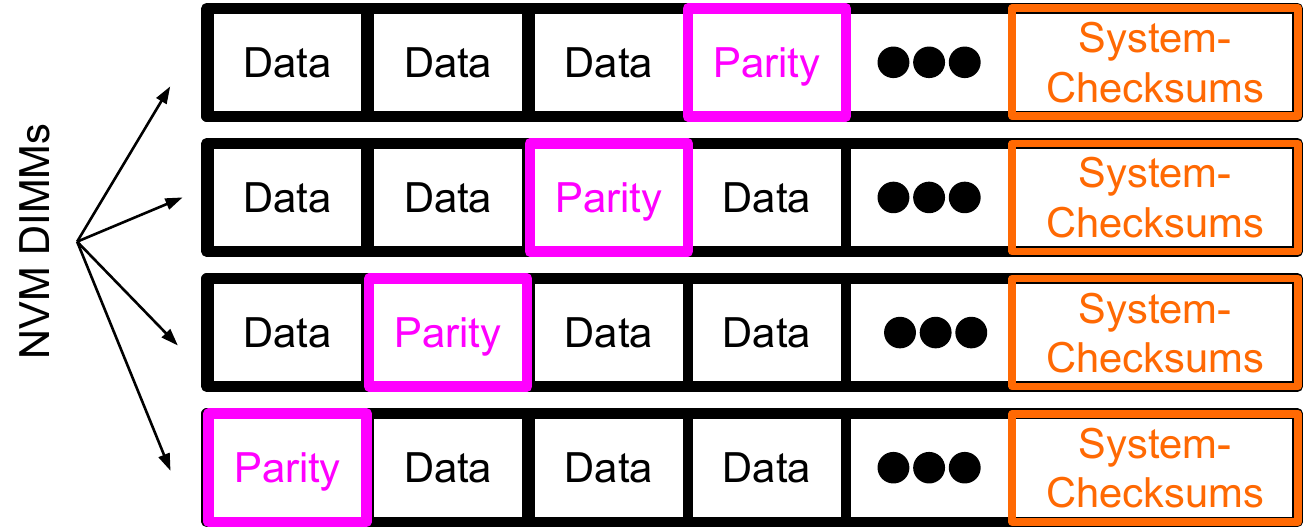}}
	\caption{\system coordinates with the file system to maintain
	per-page system-checksums and cross-DIMM parity akin 
	to RAID-5 with page striping.}
	\label{fig:redundancy-mechanisms}
\end{figure}

\system intends to enable the following for DAX-mapped 
NVM data: (i) detection of firmware-bug induced data 
corruption, (ii) recovery from such corruptions. 
To this end, the file system and \system maintain per-page 
system-checksums and cross-DIMM parity with page striping, 
as shown in \cref{fig:redundancy-mechanisms}. 

\system's redundancy mechanisms co-exist 
with other complementary file system redundancy 
mechanisms that each serve a different purpose.
These complementary mechanisms do not protect 
against firmware-bug-induced corruptions, and 
\system does not intend to protect against 
the failures that these mechanisms cover. 
Examples of such complementary redundancy mechanisms 
include remote replication for machine 
failures~\cite{designing-for-disasters-fast, 
seneca-atc, snapmirror-fast, gfs-sosp}, 
snapshots for user errors~\cite{nova-fortis-sosp, 
zfs-end-to-end-integrity-fast, btrfs-tos, wafl-usenix}, 
and inline sanity checks for file 
system bugs~\cite{wafl-integrity-fast}. 

Although not \system's 
primary intent, \system also aids in protecting 
data from random bit flips and in recovery from 
DIMM failures. 
\system can detect random bit flips because 
it maintains a checksum over the data. This coverage 
is in concert with device-level 
ECCs~\cite{chipkill-whitepaper, lot-ecc-isca, chipkill-arcc-hpca} 
that are designed for detecting and recovering from random bit flips.
\system's cross-DIMM parity also enables recovery from 
DIMM failures. The file system and \system ensure that 
recovery from a DIMM failure does not use corrupted data/parity from other 
DIMMs.
To that end, the system-checksum for a page is stored in the 
same DIMM as the page, and the file system verifies a 
page's data with its system-checksum
before using it.

\newcommand{\naive}{\textsc{Naive}\xspace}
\newcommand{\fverify}{\textsc{EV}\xspace}
\newcommand{\fverifyupdate}{\textsc{EVU}\xspace}

\subsection{Basic Redundancy Controller Design}
\label{sec:basic-design}
\begin{figure}[t]
	\centering
	\vspace*{10mm}
	{\includegraphics[width=0.8\columnwidth]{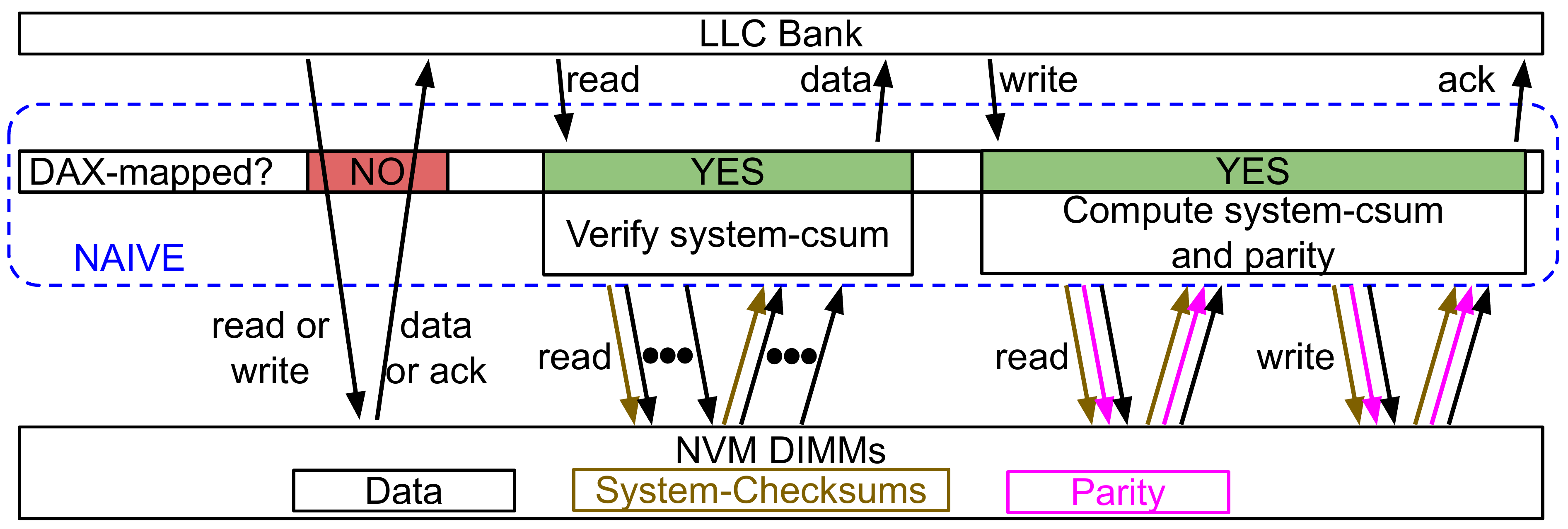}}
	\caption{\textbf{Basic Design}:
		\naive operates only 
	on DAX-mapped data. For DAX-mapped cache-line reads, \naive
	reads the entire page to compute the checksum, reads the 
	system-checksum, and verifies that the two match.
	For cache-line writes, \naive
	reads the old data, system-checksum, and parity, computes 
	the data diff, uses that to compute the new system-checksum 
	and parity, and writes them back to NVM.}
	\label{fig:basic-design}
\end{figure}

\cref{fig:basic-design} illustrates a basic redundancy 
controller design that satisfies 
the requirements for 
detecting firmware-bug induced corruptions, as described in 
\cref{sec:redundancy-mechanisms}. We refer 
to this basic design as \naive, and will 
improve \naive's design to build up to \system.
\naive resides 
above the device firmware in the data path (with 
the LLC bank controllers). 
The file system informs \naive about
physical page ranges of a file when it DAX-maps 
the file, along with the corresponding 
system-checksum pages and the parity scheme. 
For each cache-line write-back from the 
LLC and cache-line read into the LLC, 
\naive performs an address range matching. 
\naive does not do anything for cache-lines 
that do not belong to a DAX-mapped regions, 
as illustrated in the leftmost read/write access 
in \cref{fig:basic-design}.
The file system continues 
to maintain the redundancy for such 
data~\cite{nova-fortis-sosp, plexistore-nvmw}.

For DAX-mapped cache-lines, 
\naive updates and verifies  redundancy 
using separate accesses 
from the corresponding data.
The request in the center of \cref{fig:basic-design} 
shows a DAX cache-line read. 
To verify the read, 
\naive reads the entire page (shown with 
black arrows), computes 
the page's checksum, 
reads the page's system-checksum (shown in olive) 
and verifies that the two match. 
The rightmost request in 
\cref{fig:basic-design} shows a cache-line write. 
\naive reads the 
old data in the cache-line, the old
system-checksum, and the old parity (illustrated using black, 
olive and pink, respectively). It then computes the 
data diff using the old and the new data  
and uses that to compute the new system-checksum and 
parity values\footnote{We assume
	that the storage system implements 
	incremental system-checksums that 
	can be updated using the data diffs, e.g., CRC.}. 
It then writes the new data, new system-checksum, and the 
new parity to NVM. 
\naive's cross-DIMM parity design and the use 
of data diffs to update parity 
is similar to recently proposed RAIM-5b~\cite{raim-isca}. 

\naive, and consequently \system, assume that the storage servers are 
equipped with backup power to 
flush CPU caches in case of a power failure. 
The backup power guarantees that 
\naive and \system can complete 
the system-checksum and 
parity writes corresponding to a data write in case 
of a power failure, even if they cache this 
information, as we will describe later. 
This backup power could 
either be from top-of-the-rack batteries with 
OS/BIOS support to flush caches,
or ADR-like support for caches with 
cache-controller managed flushing. 
Both of these designs are common in production 
systems~\cite{viyojit-isca, no-compromises-sosp, 
wafl-usenix, wsp-asplos, 
pcommit-deprecated, kiln-micro, agigaram-nvdimm}.
Backup power also eliminates the need for 
durability-induced cache-line flushes and improves 
performance~\cite{kiln-micro, wsp-asplos}. 
We extend this assumption, and the corresponding 
performance benefits, to the all the designs 
we compare \system to in \cref{sec:eval}. 

\subsection{Efficient Checksum Verification}
\label{sec:tcsums}
Verifying system-checksums in \naive 
incurs a high overhead 
because it has to read the entire page, 
as shown in \cref{fig:basic-design}.
For typical granularities of 
4KB checksum pages and and 64B cache lines, 
\naive reads 65$\times$ more cache lines 
(64 cache-line in a page and one for the checksum).
Although a smaller checksum granularity would
reduce the checksum verification overhead,
doing so would require dedicating more of 
the expensive NVM storage for redundant data. 
For example, per-cache-line checksums would
require 64$\times$ more space than per-page checksums.
Indeed, the trend in storage system designs is 
to move towards  
larger, rather than smaller, checksum 
granularities~\cite{linux-4k-sectors, toshiba-4k-sectors, wd-4k-sectors}.

\begin{figure}[t]
	\centering
	\vspace*{10mm}
	{\includegraphics[width=0.8\columnwidth]{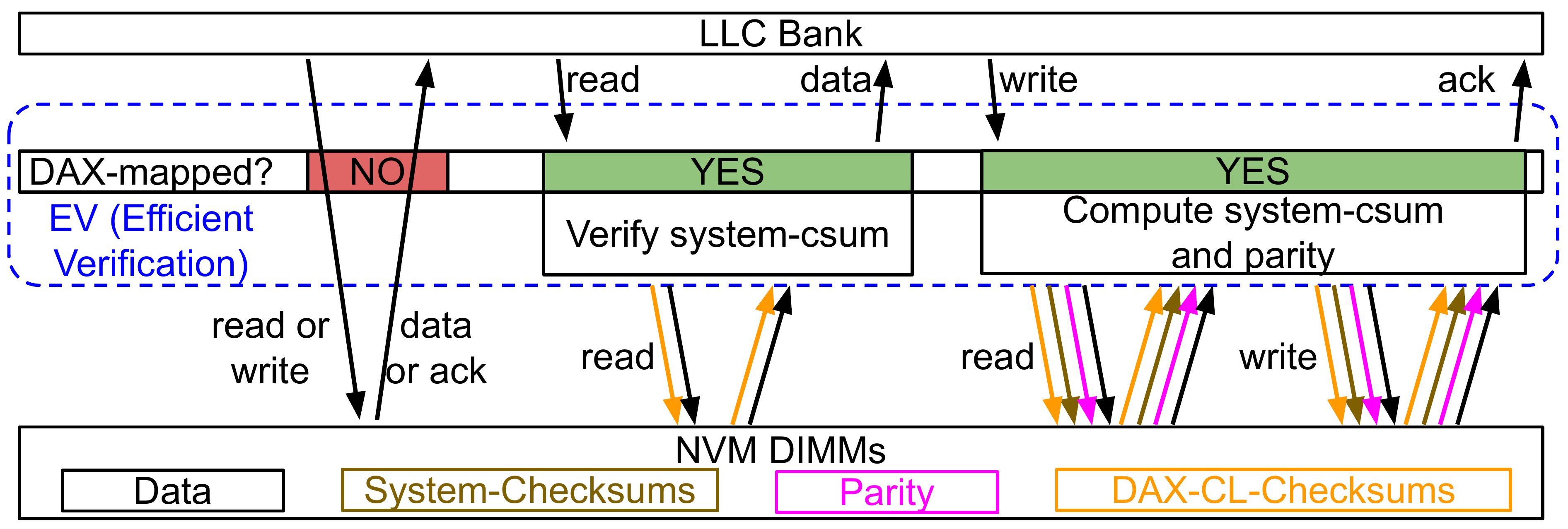}}
	\caption{\textbf{Efficient Checksum Verification}:
		\tcsums eliminate the 
		need to read the entire page for DAX  
		cache-line read verification. Instead, \fverify 
		only reads the cache-line and its corresponding 
		\tcsum.}
	\label{fig:cl-csums}
\end{figure}

We introduce \emph{\tcsums} to  
reconcile the performance overhead of 
page-granular checksum 
verification with the space overhead of cache-line checksums.
Adding \tcsums to \naive results in the 
\fverify (Efficient Verification) design 
shown in \cref{fig:cl-csums}.
As the name suggests, \tcsums 
are cache-line granular checksums that 
\fverify maintains only when data is DAX-mapped. 
The read request in the middle of 
\cref{fig:cl-csums} illustrates that using 
\tcsums reduces the read amplification 
to only 2$\times$ from 
65$\times$ for \naive---\fverify  
only needs to read the \tcsum in addition to 
the data to verify the read. 
The additional space for \tcsums  
is required only for 
the fraction of NVM that is DAX-mapped, in contrast to 
maintaining cache-line-granular or object-granular checksums for 
all NVM data at all times~\cite{pangolin-atc}.

\fverify accesses \tcsums separately from
the corresponding data to ensure that the it 
continues to provide protection from firmware-bug-induced 
corruptions. For DAX-mapped cache-line writes, 
\fverify updates the corresponding \tcsum as well, 
using a similar process as that for system-checksums 
and parity (rightmost request in \cref{fig:cl-csums}). 

Managing \tcsums is simple 
because \fverify uses them only 
while data is DAX-mapped. 
In particular, 
when recovering from any failure or crash, the file system verifies 
data integrity using system-checksums rather than \tcsums.
Thus the file system and \fverify can afford to lose 
\tcsums in case of a failure. When the file system DAX-maps a file, 
\fverify requests a buffer space for 
\tcsums. The file system can allocate this buffer space in 
either NVM or in DRAM; our implementation stores \tcsums in NVM. 
The file system reclaims this space when it 
unmaps a file. Unlike page system-checksums, 
\tcsums need not be stored on the 
same DIMM as its corresponding data because they are 
not used to verify data during recovery from a DIMM failure. 

\subsection{Efficient Checksum and Parity Updates}
\label{sec:redundancy-caching}
The rightmost request in \cref{fig:cl-csums} 
shows that \fverify incurs 4 extra NVM reads and writes 
for each cache-line write to update the 
redundancy. 
To reduce these NVM accesses, 
we note that redundancy information is 
cache-friendly. 
Checksums are typically small and multiple 
checksums fit in one cache line. 
In our implementation of 4 byte CRC-32C checksums, 
one 64 byte cache-line holds
16 checksums. \Tcsums for consecutive cache-lines 
and system-checksums for consecutive physical pages  
in a DIMM belong to the same cache-line. 
Access locality in data 
leads to reuse of \tcsum and system-checksum cache-lines.
Similarly, accesses to logically consecutive pages 
lead reuse of parity cache-lines because 
they belong to the same RAID stripe. 

\begin{figure}[t]
	\centering
	\vspace*{10mm}
	{\includegraphics[width=0.8\columnwidth]{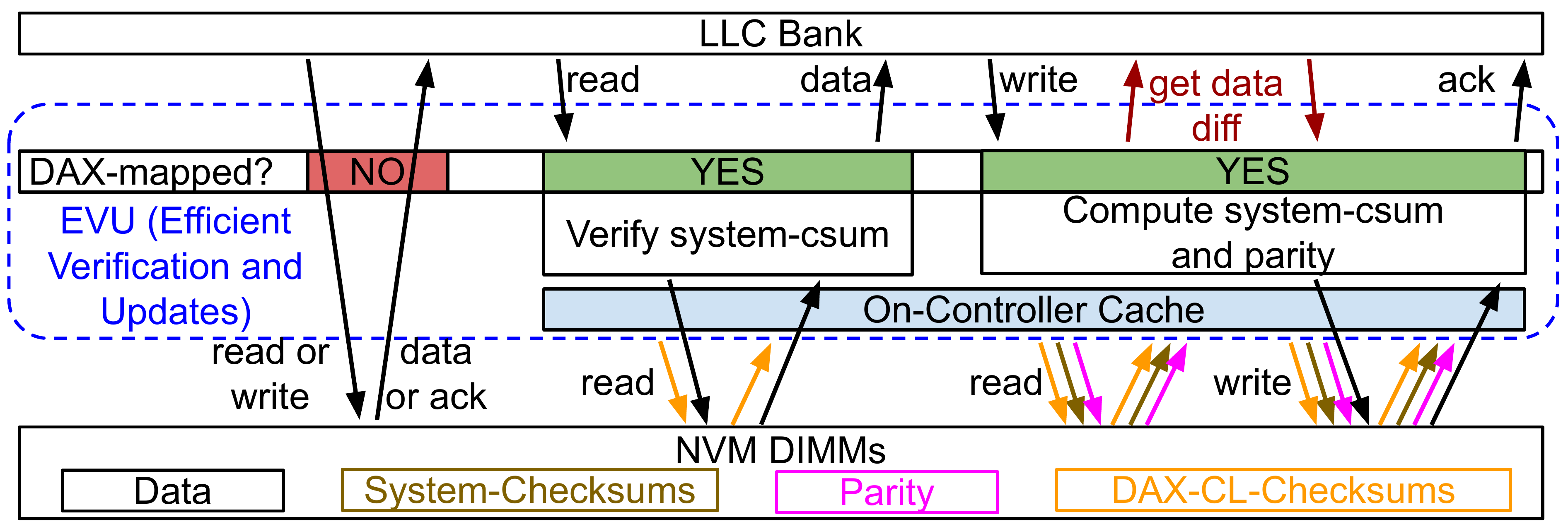}}
	\caption{\textbf{Efficient Checksum and Parity Updates}:
		\fverifyupdate caches redundancy cache-line in an on-controller 
		cache and a LLC partition (not shown). 
		\fverifyupdate also uses a LLC partition to store data diffs,  
		eliminating the need to read the old 
		data from NVM upon cache-line write-backs.}
	\label{fig:llc-caching}
\end{figure}

\cref{fig:llc-caching} shows \fverifyupdate (Efficient Verification and Updates) 
that, in addition 
to \fverify, caches the redundancy data, i.e., system-checksums, 
\tcsums, and parity, in a small on-controller cache.
\fverifyupdate does not cache the corresponding 
NVM data because the LLC already does that. 
\fverifyupdate also uses a partition of the LLC  
to increase its cache space for redundancy 
information (not shown in the figure).  
Using a reserved LLC partition for caching 
redundancy information limits the interference 
with application data. 
\fverifyupdate can insert up to 3 
redundancy cache-lines 
(checksum, \tcsum, and parity) per data 
cache-line write-back. 
If \fverifyupdate were to share the entire LLC for 
caching redundancy information, each of these redundancy 
cache-lines could force out application 
data. 
Reserving a partition for redundancy information eliminates 
this possibility because 
\fverifyupdate can only evict a redundancy cache-line when 
inserting a new one. 

\fverifyupdate also eliminates the need to fetch the 
old data from NVM to compute the data diff. 
Cache-lines in the LLC become dirty when they are evicted 
from the L2. Since the LLC already contains 
the soon-to-be-old data value, 
\fverifyupdate uses it to compute the data diff and 
stores this diff in a LLC partition. 
This enables \fverifyupdate to directly use this data diff 
upon a LLC cache-line write back (shown as maroon 
arrows from EVU to LLC in the rightmost request in 
\cref{fig:llc-caching}). 
Upon an eviction 
from the LLC data diff partition (e.g., to insert 
a new data diff), \fverifyupdate writes-back 
the corresponding data without evicting 
it from the LLC, and marks the data 
cache-line as clean in the LLC.
This ensures that the future eviction of the 
data cache-line would not require \fverifyupdate 
to read the old data either, while 
allowing for reuse of the data in the LLC. 

\fverifyupdate's LLC partitions 
(for caching redundancy and storing data diffs) 
are completely decoupled from the application data 
partitions. The cache controllers do not 
lookup application data in \fverifyupdate's partitions, and 
\fverifyupdate does not look up redundancy 
or data diff cache-lines in application data partitions. 

\subsection{Putting it all together with \system}
\label{sec:architecture}
\begin{figure}[t]
	\centering
	{\includegraphics[width=0.8\columnwidth]{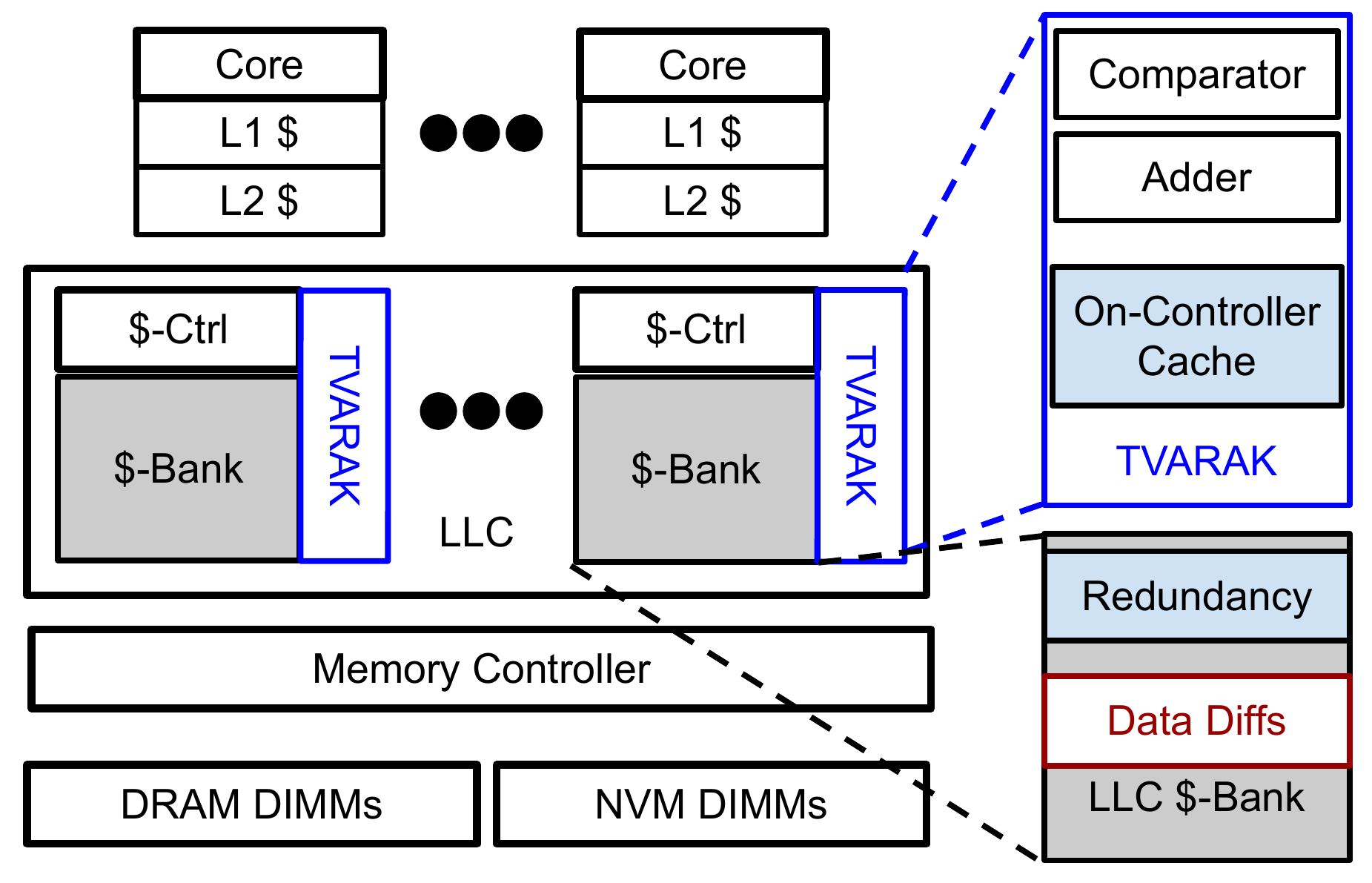}}
\caption{\system is co-resides 
	with the LLC bank controllers. 
	It includes comparators to  
	identify cache-line that belong to DAX-mapped pages 
	and adders to compute checksums and parity. It includes a small 
	on-controller redundancy cache that is backed by a 
	LLC partition. \system also stores the data diffs 
	to compute checksums and parity.}
\label{fig:hw-components}
\end{figure}

\cref{fig:hw-components} shows \system's components, 
which are based on \fverifyupdate's design.
One \system controller co-resides
with each LLC cache bank. 
Each \system controller consists of comparators 
for address range matching and adders for checksum and 
parity computations. 
\system includes a 
small on-controller cache for  
redundancy data and uses LLC 
way-partitions for caching redundancy data 
and storing data diffs. 
The controllers use MESI coherence protocol for sharing the 
redundancy cache-lines between their private caches.  

\textbf{Area Overhead}: The on-controller cache 
dominates \system's area overhead because its other 
components (comparators and adders) only require 
small logic units. 
In our evaluation with 2MB LLC cache banks, each \system controller 
consists of a 4KB cache. 
This implies that \system's area is only 0.2\% of the LLC.
\system's design of caching redundancy in an 
LLC partition instead of using its own large cache 
keeps \system's dedicated area overheads low, 
without compromising on performance (\cref{sec:eval}).

\textbf{Life of DAX-mapped cache-lines with \system}:
For a DAX-mapped cache-line read, \system computes the 
corresponding \tcsum address and 
looks it up in the on-controller cache. Upon a 
miss, it looks up the \tcsum in the 
LLC redundancy partition. 
If it misses in the LLC partition as well,  
\system reads the \tcsum from NVM and caches it.
\system read the data cache-line from NVM, computes its 
checksum, and verifies it with the \tcsum. 
If the checksum verification succeeds, 
\system hands over the data to the bank controller. 
In case of an error, \system raises an 
interrupt that traps into the OS; the 
file system then initiates a recovery using the 
cross-DIMM parity. 

On a DAX-mapped cache-line write, \system computes the 
corresponding system-checksum, \tcsum, and parity addresses and  
reads them following the same process as above. 
\system retrieves the data diff from the LLC bank partition
and uses that to compute the new system-checksum, \tcsum, and parity.
\system stores the updated redundancy information in the on-controller cache, 
and writes-back the data cache-line to NVM.
\system can safely cache the updated redundancy information
because it assumes that servers are equipped with 
backup power to flush caches to persistence in case of a
power failure (\cref{sec:basic-design}). 

\system fills an important gap in NVM storage 
redundancy with simple architectural changes. 
We believe that \system can be easily integrated in 
storage server chips, specially because 
integrating NVM devices into servers already 
requires changing the on-chip memory controller 
to support the new DDR-T protocol~\cite{nvm-perf-arxiv}.

%% file: eval.tex
\section{Evaluation}
\label{sec:eval}

We evaluate \system with 7 applications and with 
multiple workloads for each application. 
\cref{table:workloads} describes our applications 
and their workloads. 
Redis~\cite{redis-pmem}, 
Intel PMDK's~\cite{pmemlib} tree-based 
key-value stores (C-Tree, B-Tree, and 
RB-Tree), and N-Store~\cite{nstore-sigmod} 
are NVM applications with complex 
access patterns. 
We also use fio~\cite{fio} to 
generate synthetic sequential and random access patterns, 
and stream~\cite{stream} for 
sequential access memory-bandwidth intensive microbenchmarks.

\begin{table}[t]
    \input{tables/workloads.tex}
    \caption{Applications and their workloads.}
    \label{table:workloads}
\end{table}
\newcommand{\txbo}{TxB-Object-Csums\xspace}
\newcommand{\txbp}{TxB-Page-Csums\xspace}
We compare \system with three alternatives: 
\emph{Baseline}, \emph{TxB-Object-Csums}, and \emph{TxB-Page-Csums}.
Baseline implements no redundancy mechanisms.
TxB-Object-Csums and TxB-Page-Csums 
are software-only redundancy approaches; 
TxB-Object-Csums is based on Pangolin~\cite{pangolin-atc} 
and TxB-Page-Csums is based on Mojim~\cite{mojim-asplos} and 
HotPot~\cite{hotpot-socc}. Both \txbo and \txbp 
update system-checksums and parity when applications inform 
the interposing library after completing a write, which 
is typically at a transaction boundary (TxB). 
\txbo maintains system-checksums at an object granularity,
whereas \txbp maintains system-checksums at a page granularity.
\txbo does not need to read the entire page to compute the 
system-checksum after a write; it computes the new 
system-checksum from the new data directly. However, 
\txbo has higher space overhead because of object-granular 
checksums.
Neither \txbp nor our \txbo verify the data read 
by an application with the corresponding system-checksum;
thus, our \txbo should overpredict performance for Pangolin's
approach (which includes read verification).
\system updates system-checksums and parity upon every write-back 
from the LLC to the NVM, and verifies system-checksums 
upon every read from the NVM to the LLC. 
As mentioned in \cref{sec:basic-design}, we 
assume battery-backed CPU caches and none of 
the designs flush cache-lines for durability. 

\textbf{Methodology:} We use zsim~\cite{zsim-isca} to simulate 
a system similar to Intel Westmere processors~\cite{zsim-isca}.
\cref{table:sim-params} details our simulation parameters.
We simulate 12 OOO cores, each with 32KB private L1 
and 256KB private L2 caches. The cores share a 
24MB last level cache (LLC) with 12 banks of 2MB 
each. The simulated system consists of 6 DRAM DIMMs 
and 4 NVM DIMMs. For NVM DIMMs, we use the latency 
and energy parameters derived by Lee et al.~\cite{pcm-isca} 
(60/150 ns read/write latency, 1.6/9 nJ per read/write). 
We evaluate the impact of changing 
the number of NVM DIMMs and the underlying 
NVM technology (and the associated performance characteristics) 
in \cref{sec:sensitivity}.
We use a fixed-work methodology and perform the same amount 
of application work for each design: baseline, \system, \txbo, 
and \txbp.
Unless stated otherwise, we present the average of three runs 
for each data point with root mean square error bars. 

\begin{table}[t]
    \input{tables/sim-params.tex}
    \caption{Simulation Parameters}
    \label{table:sim-params}
\end{table}

\subsection{Key Evaluation Takeaways}
\label{sec:takeaways}
We highlight the key takeaways 
from our results before describing each 
application's results in detail. 

\begin{itemize}
	\item \system provides efficient 
		redundancy updates for application data writes, 
		e.g., with only 1.5\% overhead 
		over baseline that provides no 
		redundancy for a insert-only workload 
		with tree-based key-value stores (C-Tree, 
		B-Tree, RB-Tree).
	\item \system verifies all application data reads, 
		unlike most existing solutions, and does 
		so efficiently. For example, 
		in comparison to baseline that does not 
		verify any reads, \system 
		slows down Redis get-only workload 
		by only 3\%. 
	\item \system benefits from application data 
		access locality because that leads to 
		better cache usage for redundancy information. 
		For example, for synthetic fio benchmarks, 
		\system has negligible
		overheads with sequential accesses, but 2\% overhead for 
		random reads and 33\% for random writes, compared 
		to baseline. 
	\item \system outperforms existing software-only
		redundancy mechanisms. For example, for Nstore workloads, \txbo is 
		33--53\% slower than \system, and \txbp 
		is 180--390\% slower than \system.
	\item \system's efficiency comes without 
		an increase in (dedicated) space requirements. 
		\txbo outperforms \txbp but at the cost 
		of higher space overhead for per-object 
		checksums. \system instead 
		uses \tcsums that improve performance without 
		demanding dedicated storage. 
\end{itemize}

\begin{figure*}
	{\centering
	\framebox[0.4\textwidth][c]{
	\includegraphics[width=0.4\textwidth, keepaspectratio]{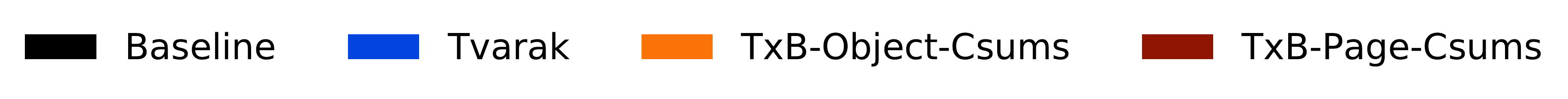}
	}
	\hfill
	\framebox[0.2\textwidth][c]{
	\includegraphics[width=0.2\textwidth, keepaspectratio]{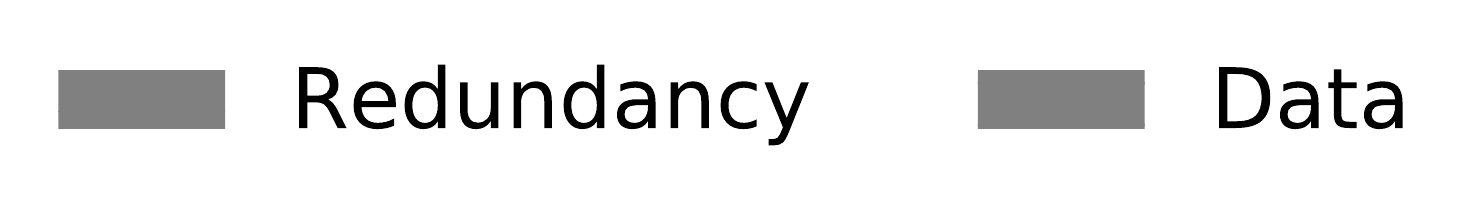}
	}	
	\hfill
	\framebox[0.3\textwidth][c]{
	\includegraphics[width=0.3\textwidth, keepaspectratio]{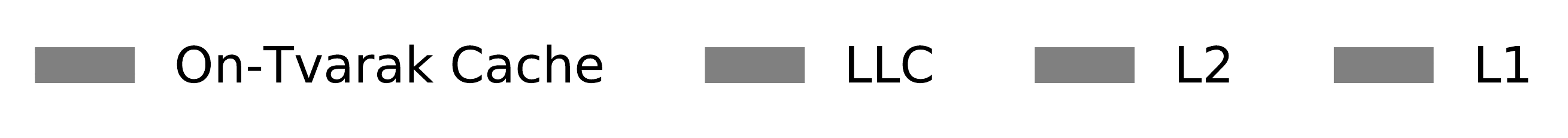}
	}
	}
	\\
	 \begin{subfigmatrix}{4}
		 \subfigure[Redis: Runtime]{\label{fig:redis-duration}
			\includegraphics[width=0.24\textwidth, keepaspectratio]{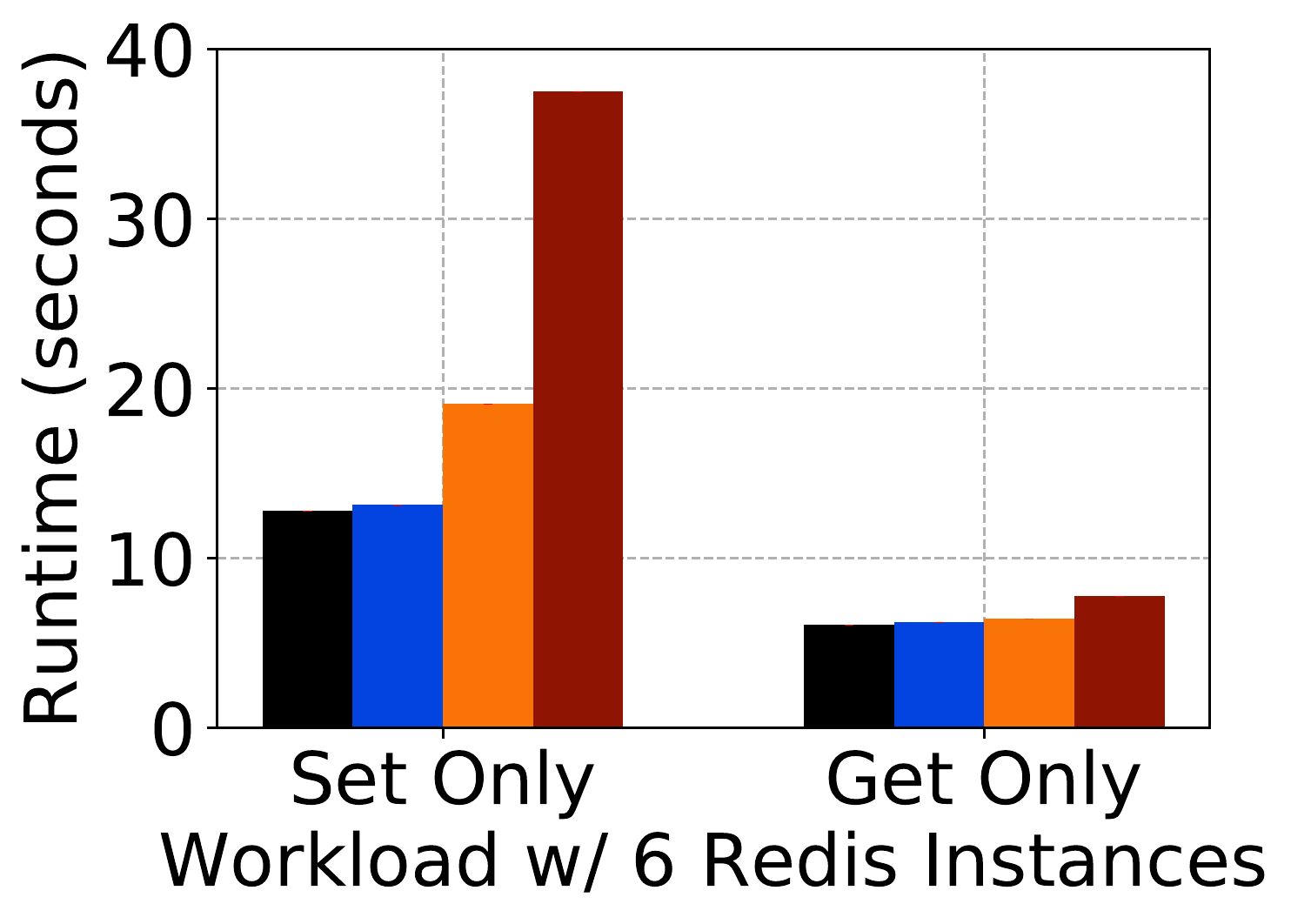}}
		\subfigure[Redis: Energy]{\label{fig:redis-energy}
			\includegraphics[width=0.24\textwidth, keepaspectratio]{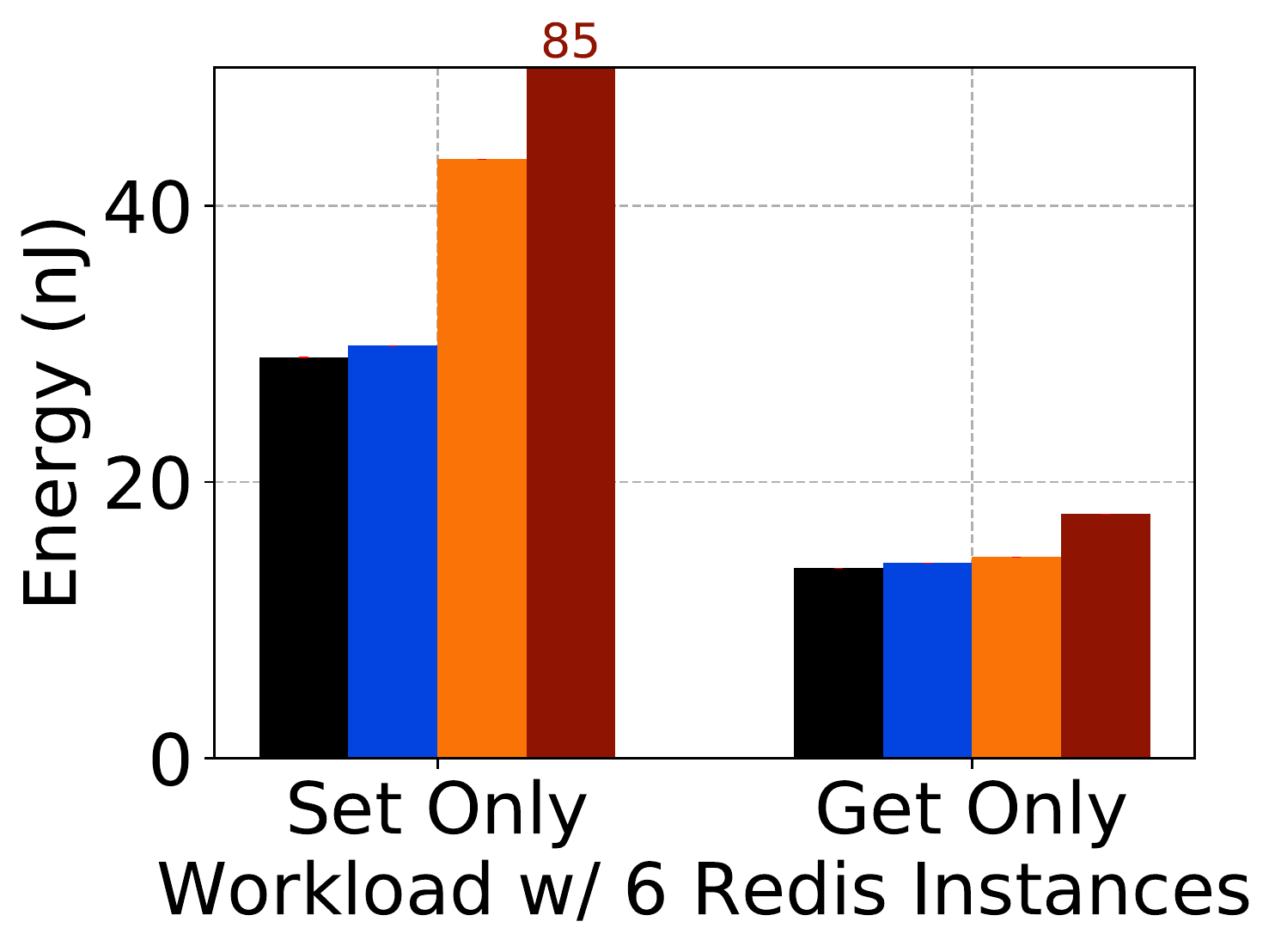}}
		\subfigure[Redis: NVM Accesses]{\label{fig:redis-nvm-accesses}
			\includegraphics[width=0.24\textwidth, keepaspectratio]{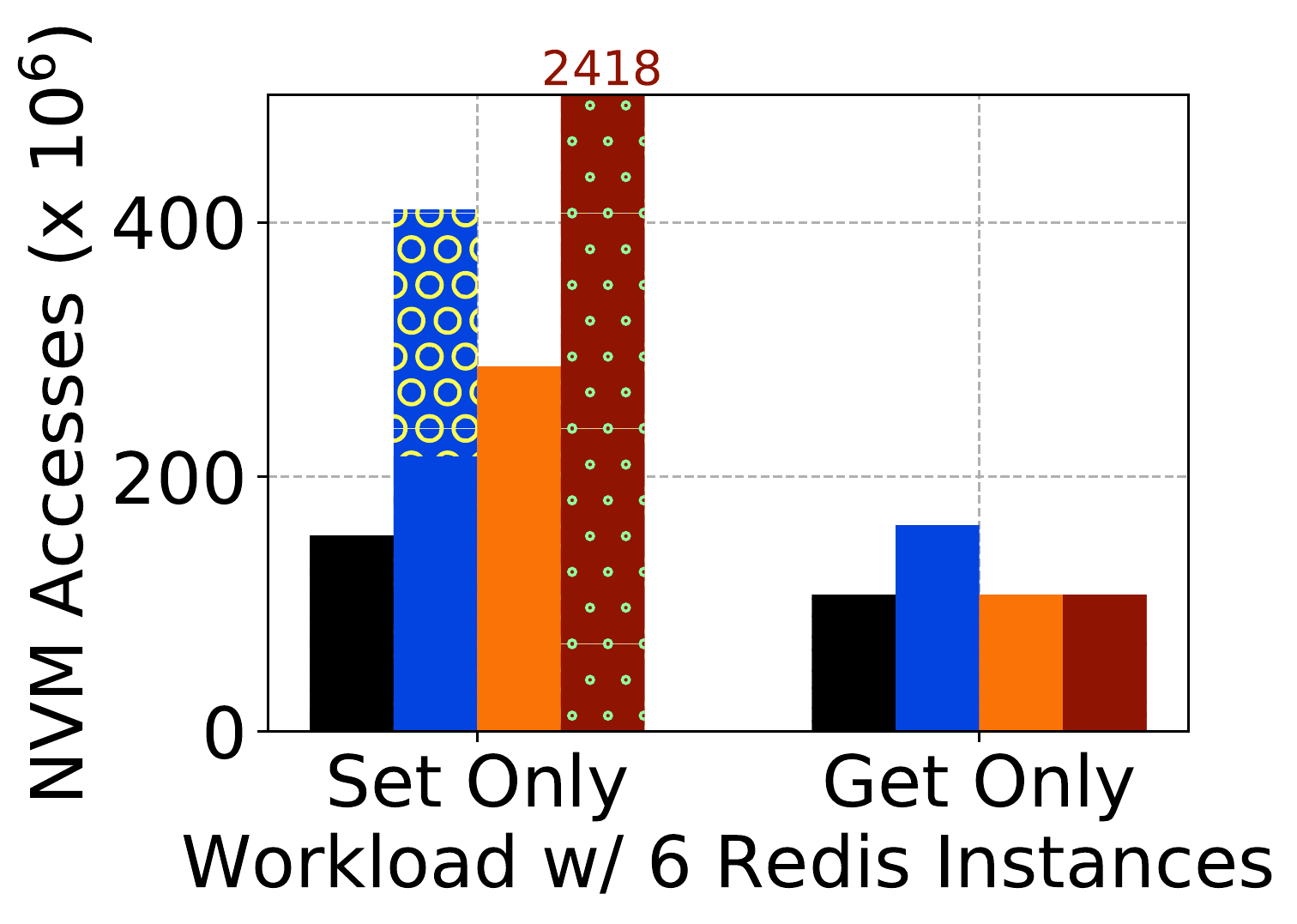}}
		\subfigure[Redis: Cache Accesses]{\label{fig:redis-cache-accesses}
			\includegraphics[width=0.24\textwidth, keepaspectratio]{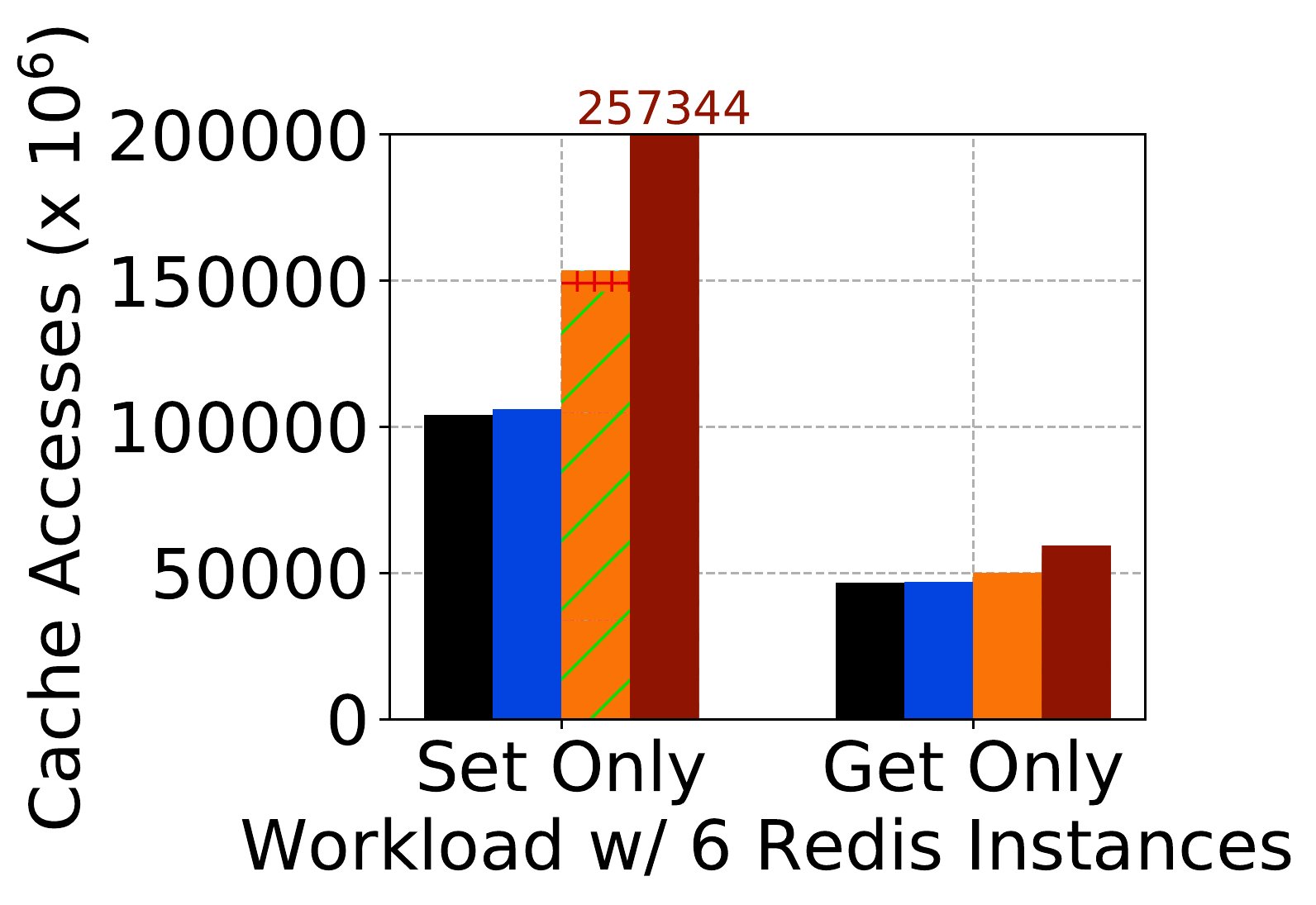}}
		 \subfigure[KV-Structures: Runtime]{\label{fig:kv-duration}
			\includegraphics[width=0.24\textwidth, keepaspectratio]{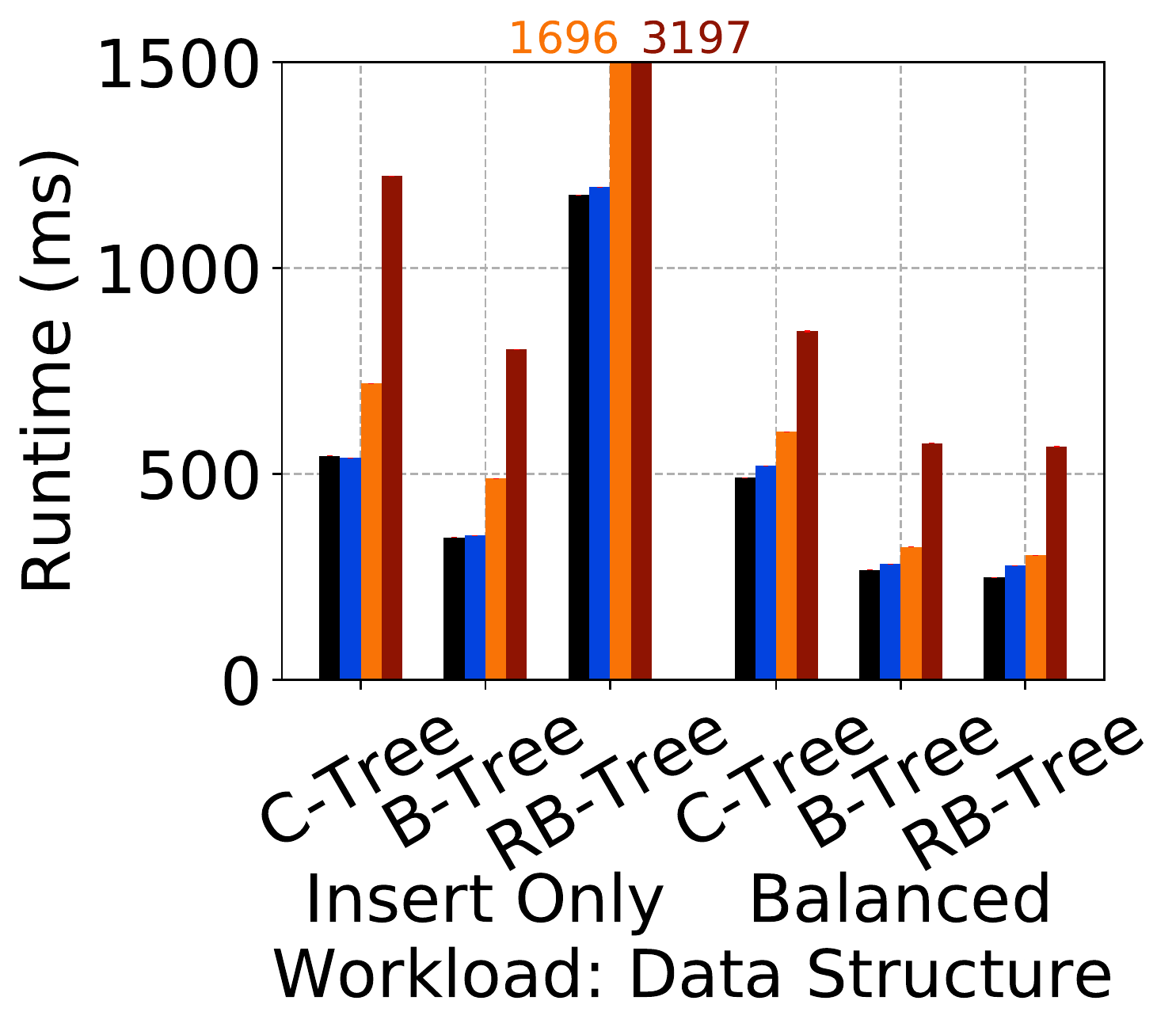}}
		\subfigure[KV-Structures: Energy]{\label{fig:kv-energy}
			\includegraphics[width=0.24\textwidth, keepaspectratio]{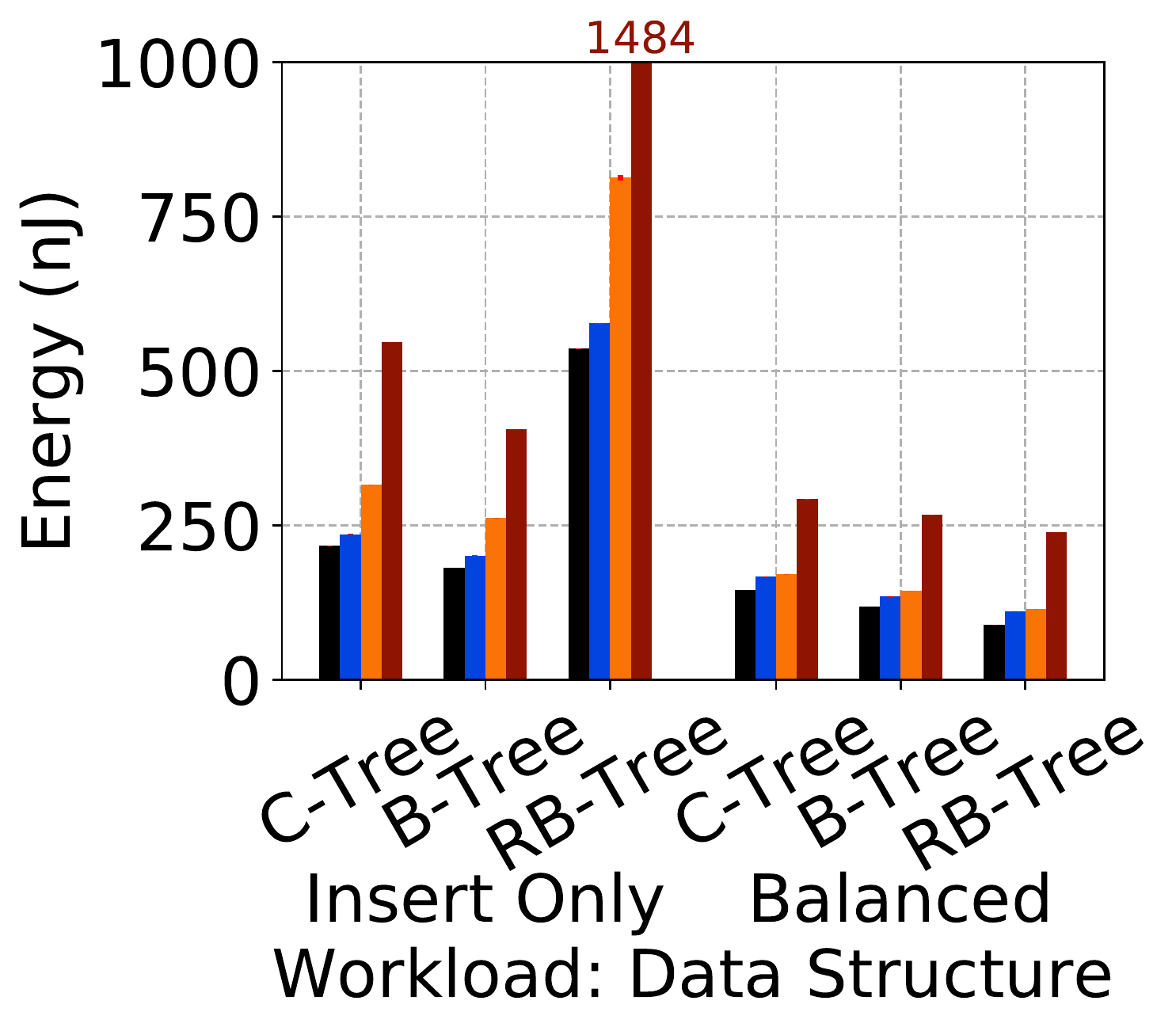}}
		\subfigure[KV-Structures: NVM Accesses]{\label{fig:kv-nvm-accesses}
			\includegraphics[width=0.24\textwidth, keepaspectratio]{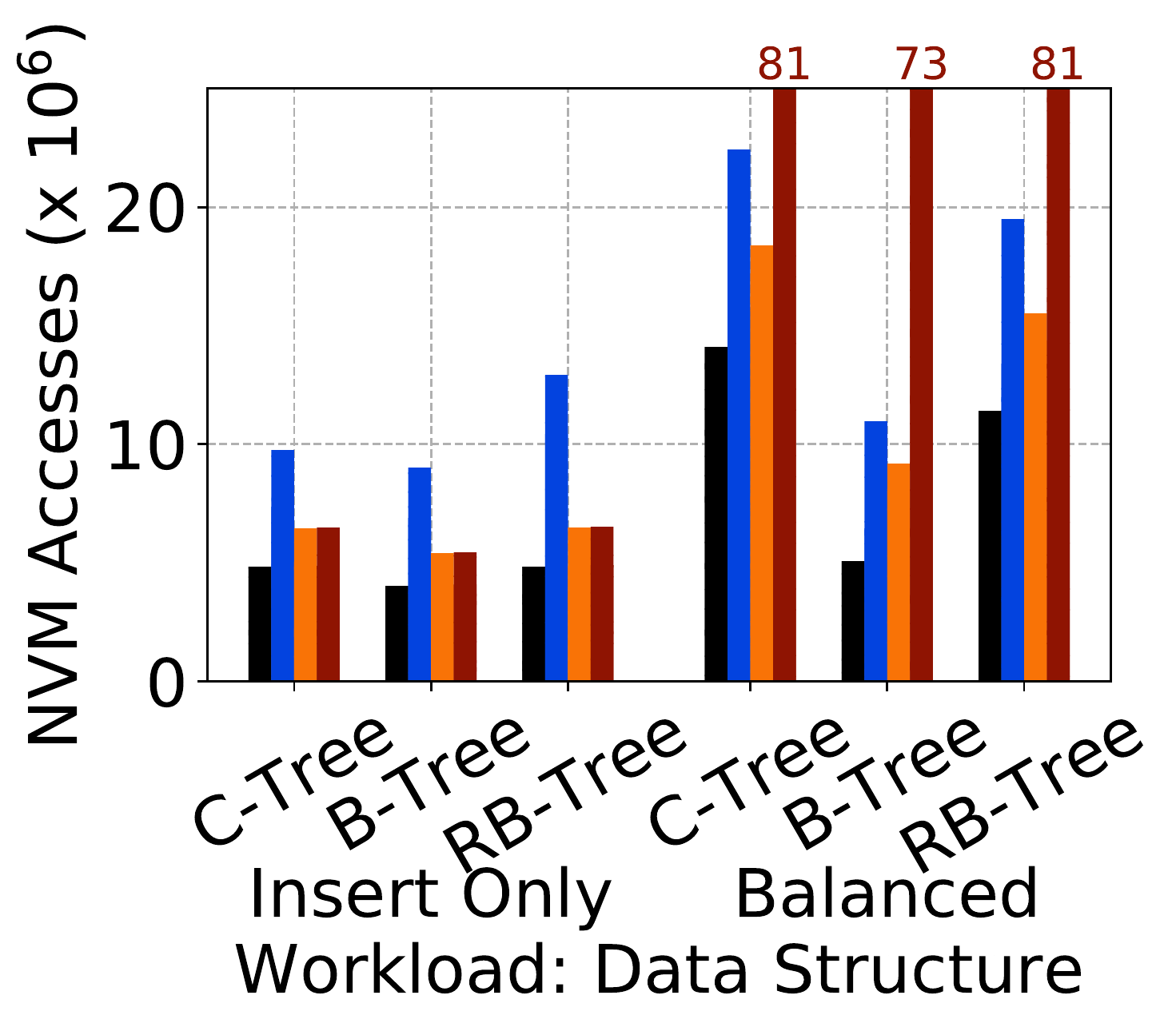}}
		\subfigure[KV-Structures: Cache Accesses]{\label{fig:kv-cache-accesses}
			\includegraphics[width=0.24\textwidth, keepaspectratio]{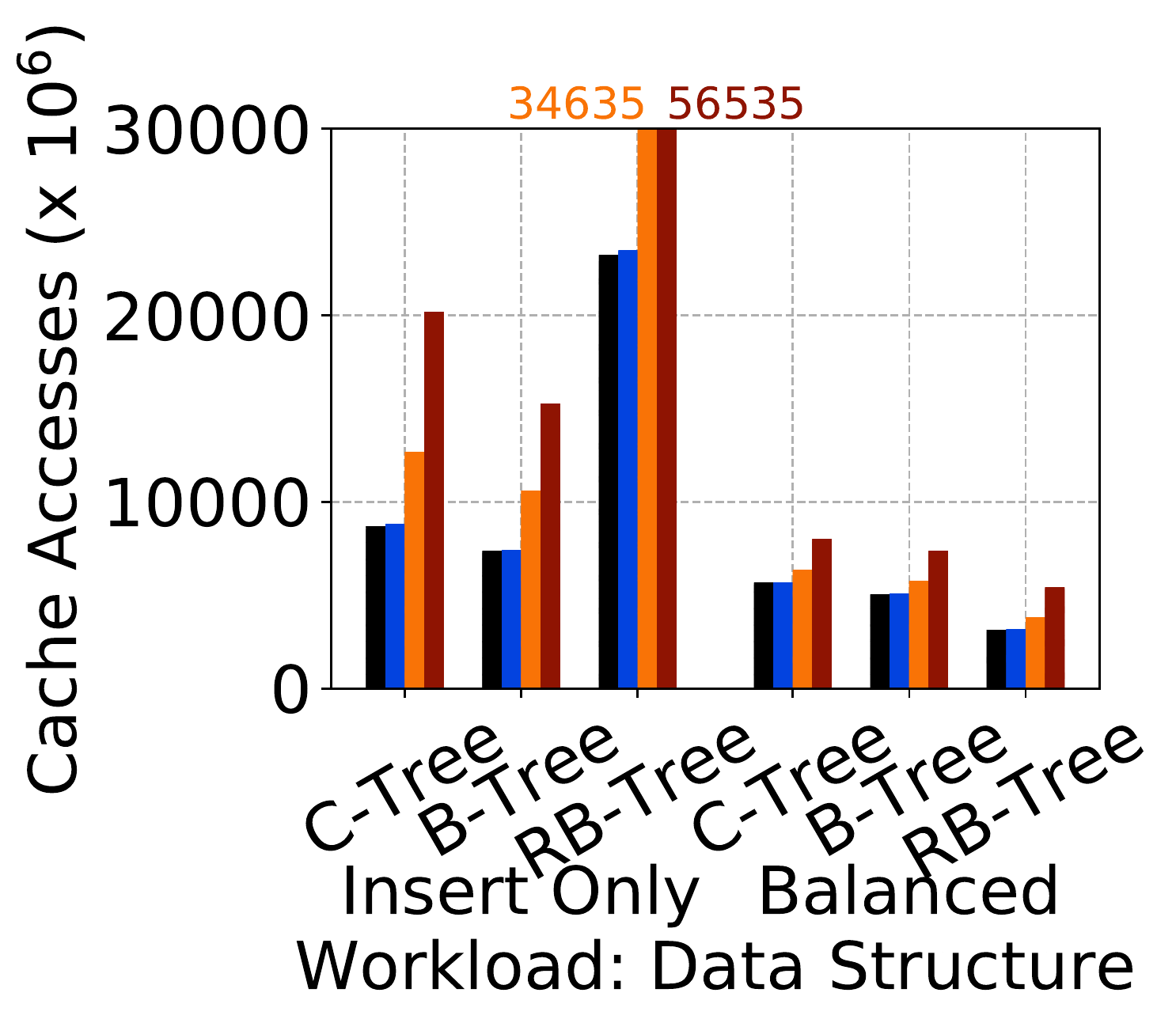}}
		\subfigure[N-Store: Runtime]{\label{fig:nstore-duration}
			\includegraphics[width=0.24\textwidth, keepaspectratio]{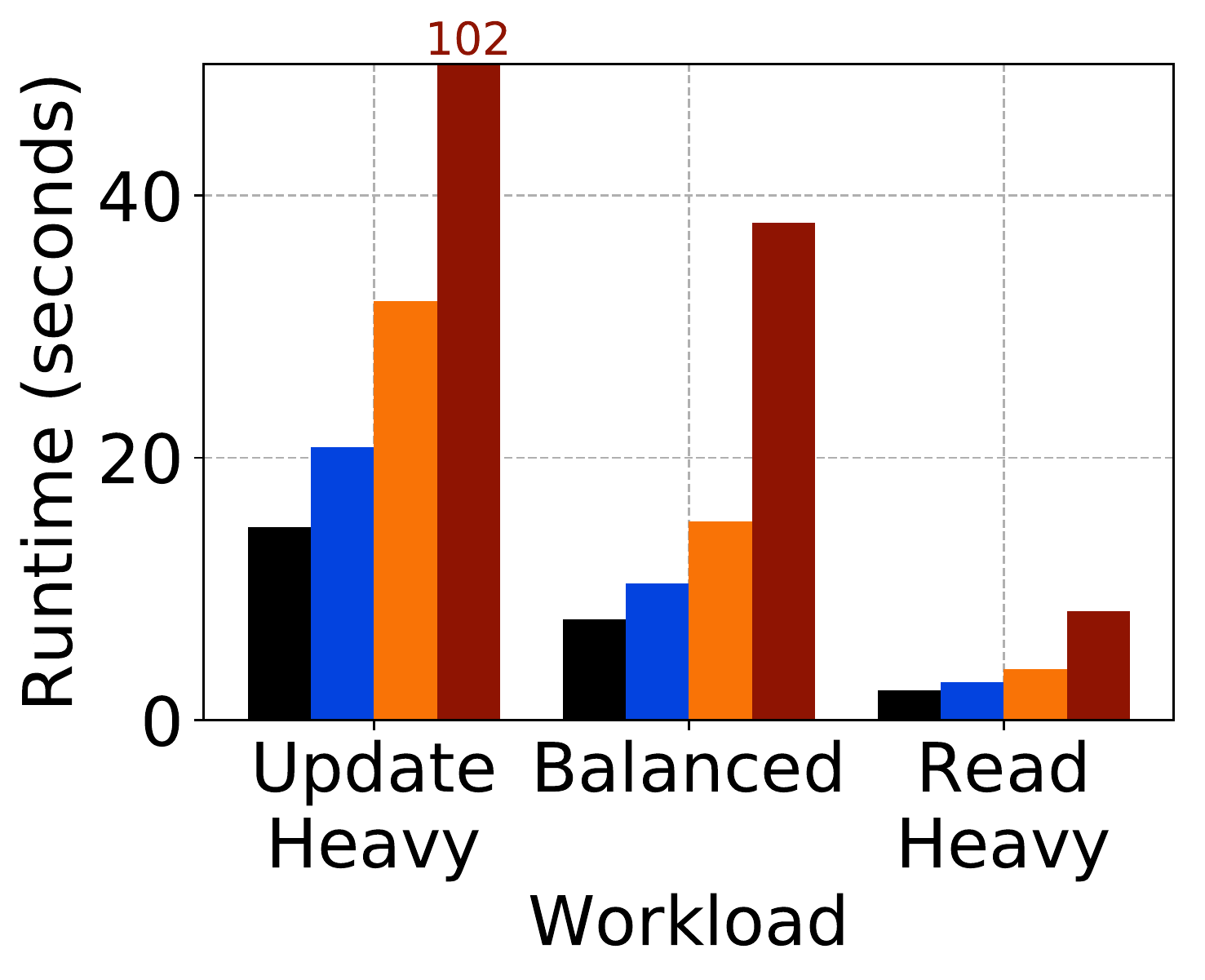}}
		\subfigure[N-Store: Energy]{\label{fig:nstore-energy}
			\includegraphics[width=0.24\textwidth, keepaspectratio]{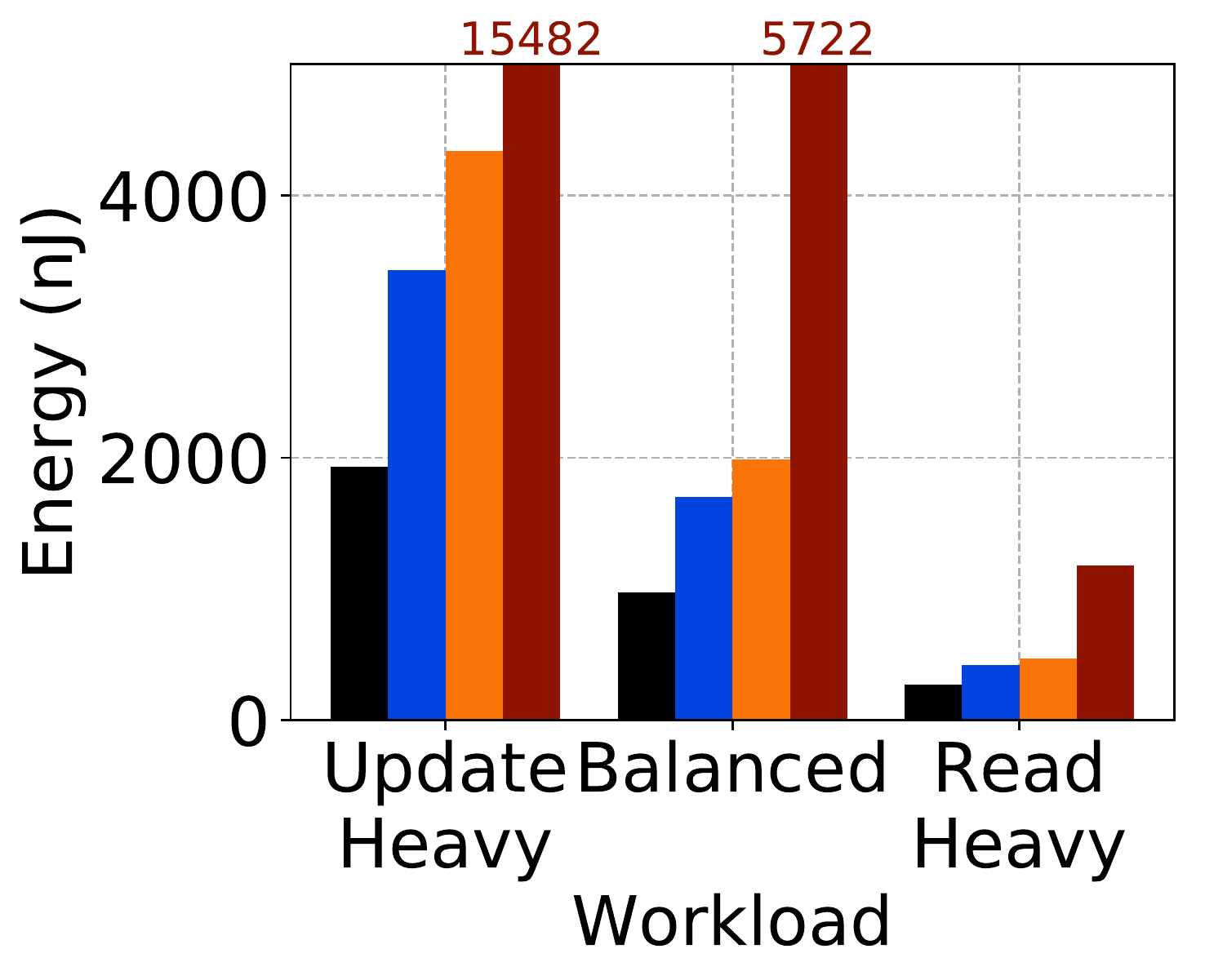}}
		\subfigure[N-Store: NVM Accesses]{\label{fig:nstore-nvm-accesses}
			\includegraphics[width=0.24\textwidth, keepaspectratio]{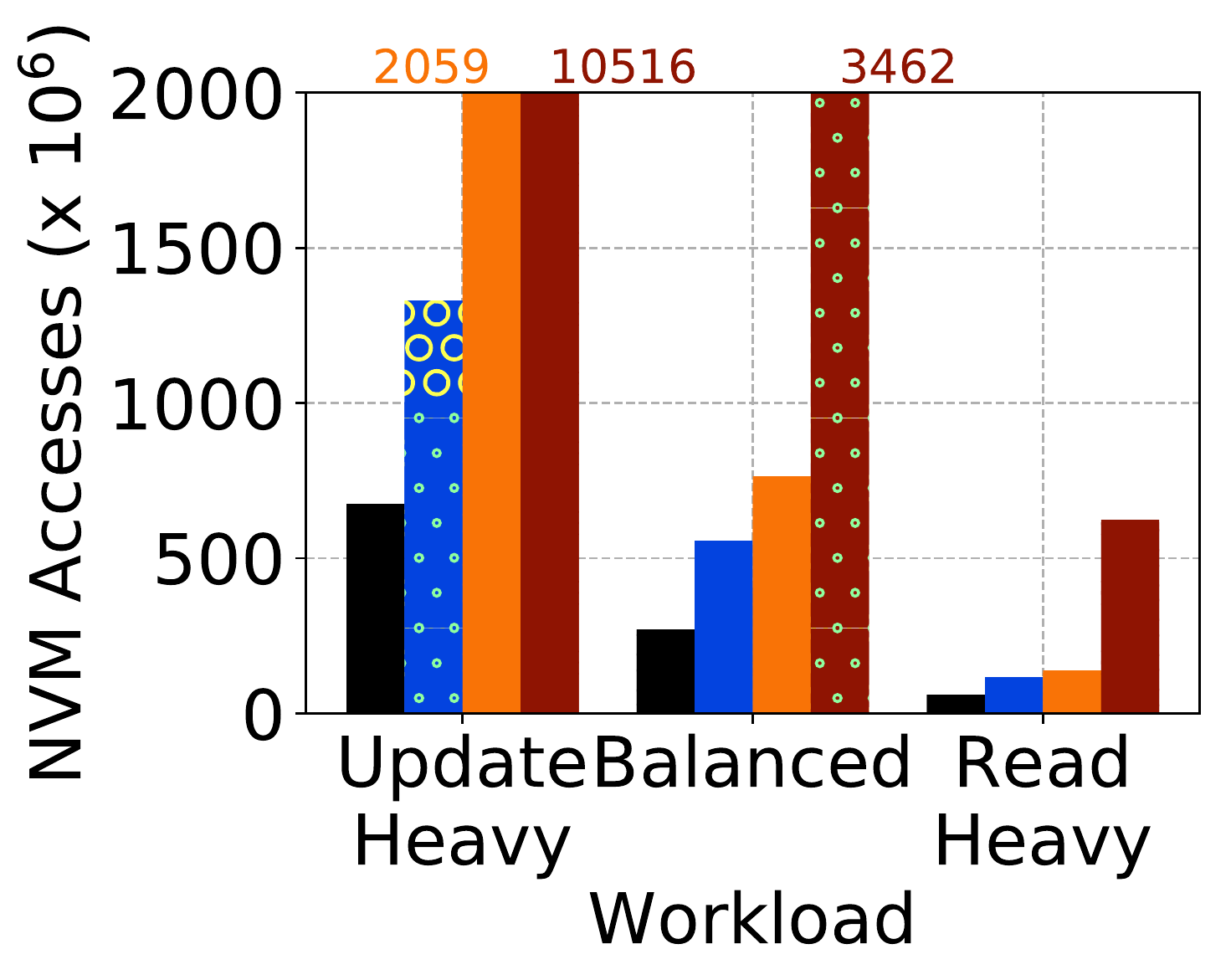}}
		\subfigure[N-Store: Cache Accesses]{\label{fig:nstore-cache-accesses}
			\includegraphics[width=0.24\textwidth, keepaspectratio]{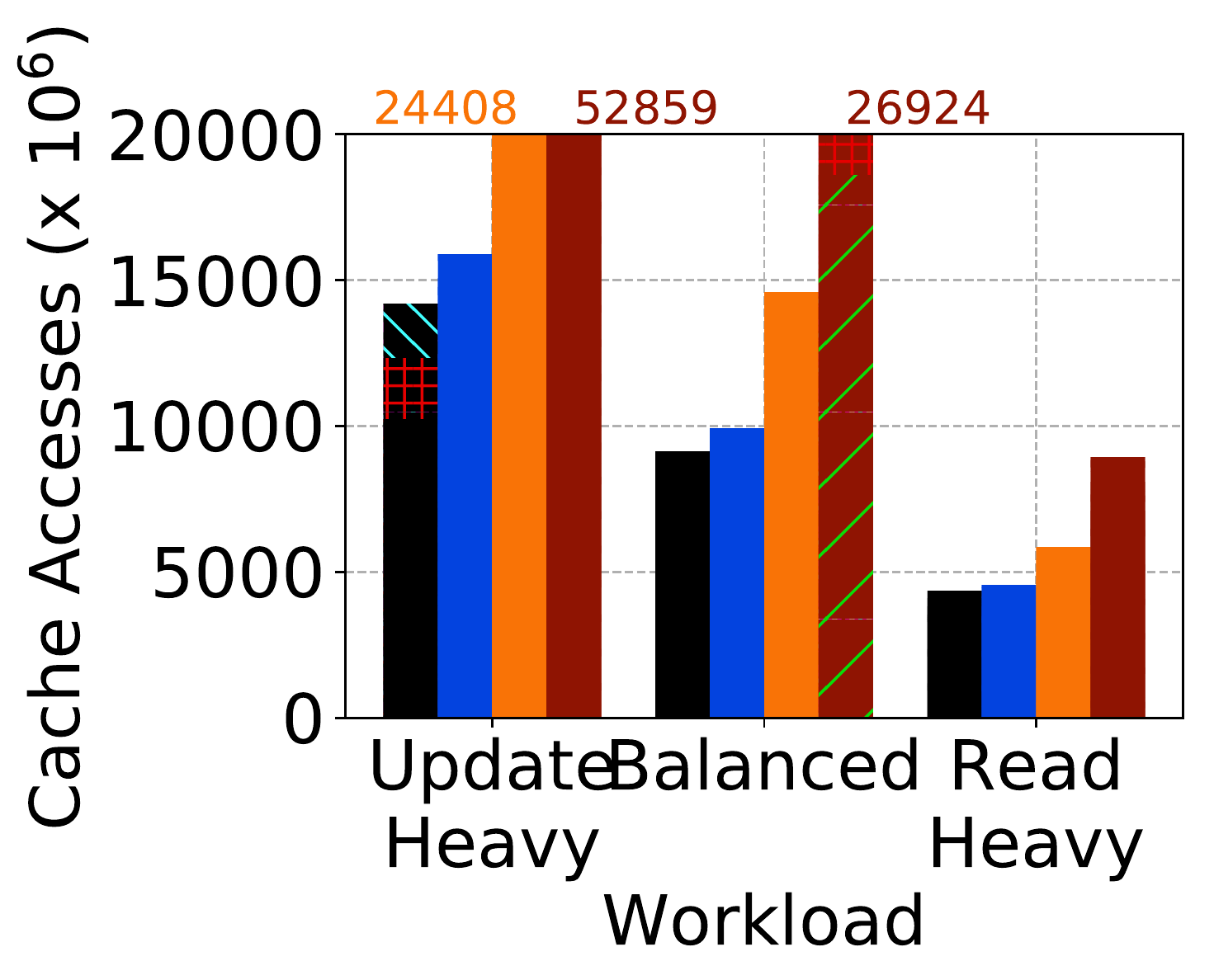}}
		\subfigure[Fio: Runtime]{\label{fig:fio-duration}
			\includegraphics[width=0.24\textwidth, keepaspectratio]{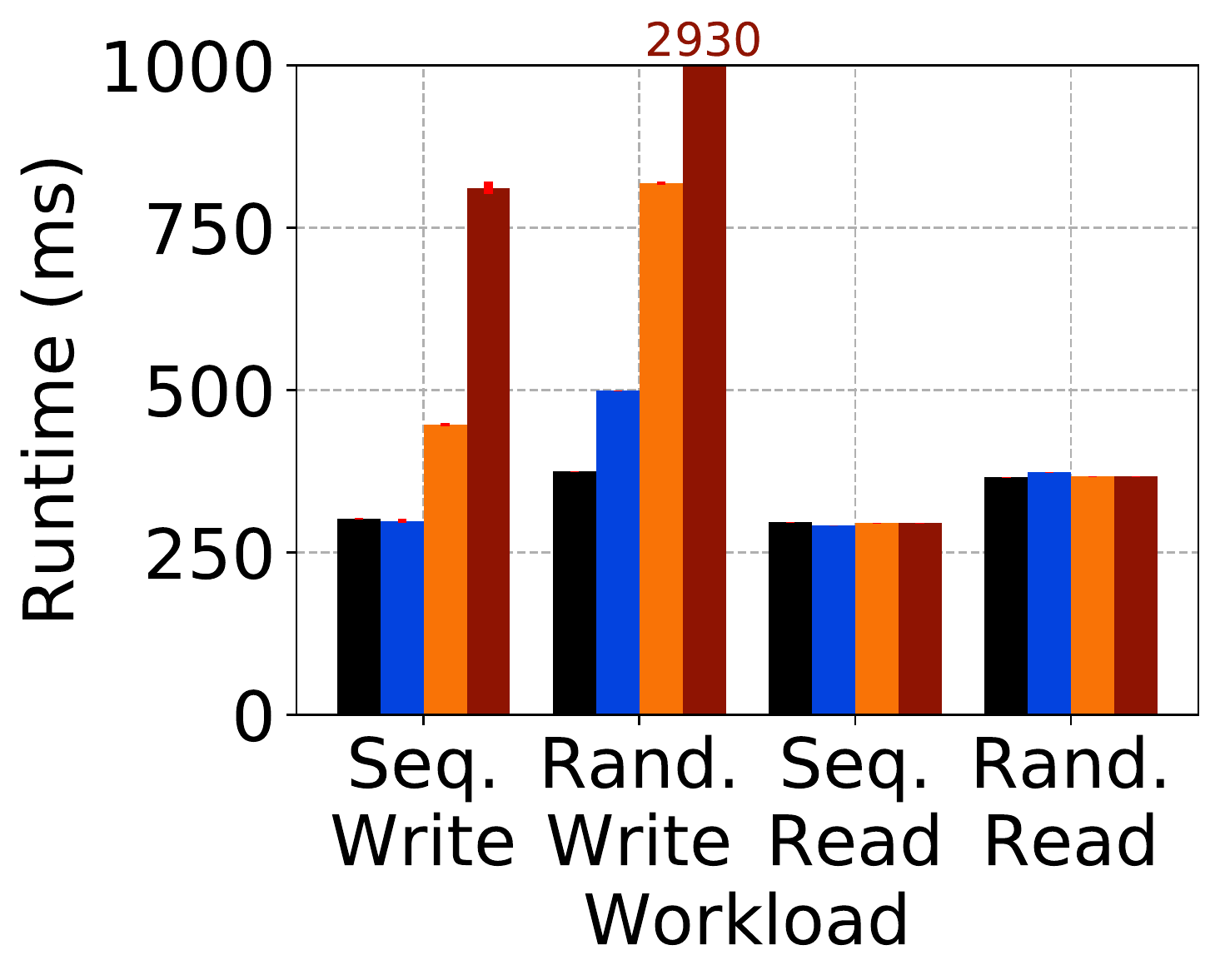}}
		\subfigure[Fio: Energy]{\label{fig:fio-energy}
			\includegraphics[width=0.24\textwidth, keepaspectratio]{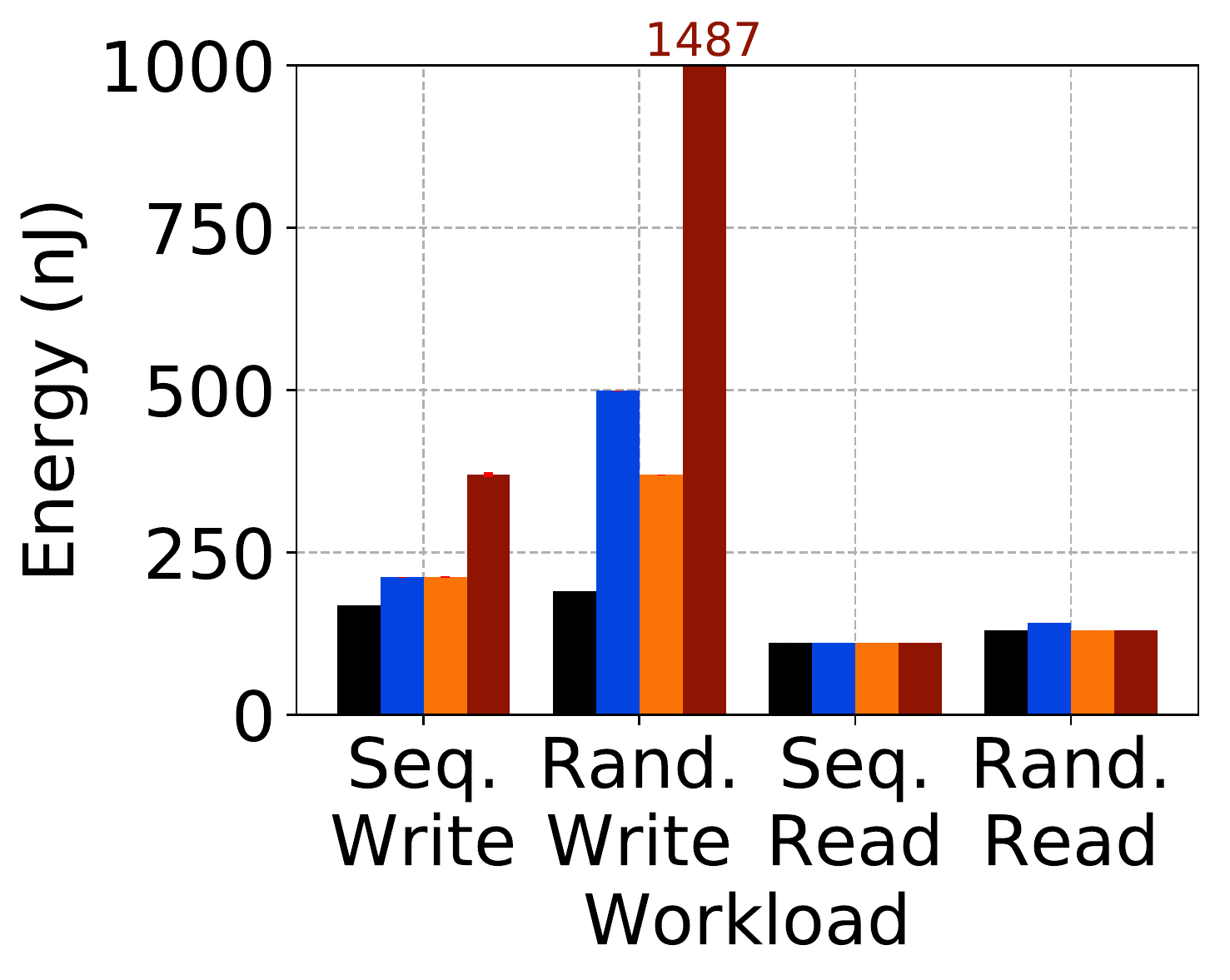}}
		\subfigure[Fio: NVM Accesses]{\label{fig:fio-nvm-accesses}
			\includegraphics[width=0.24\textwidth, keepaspectratio]{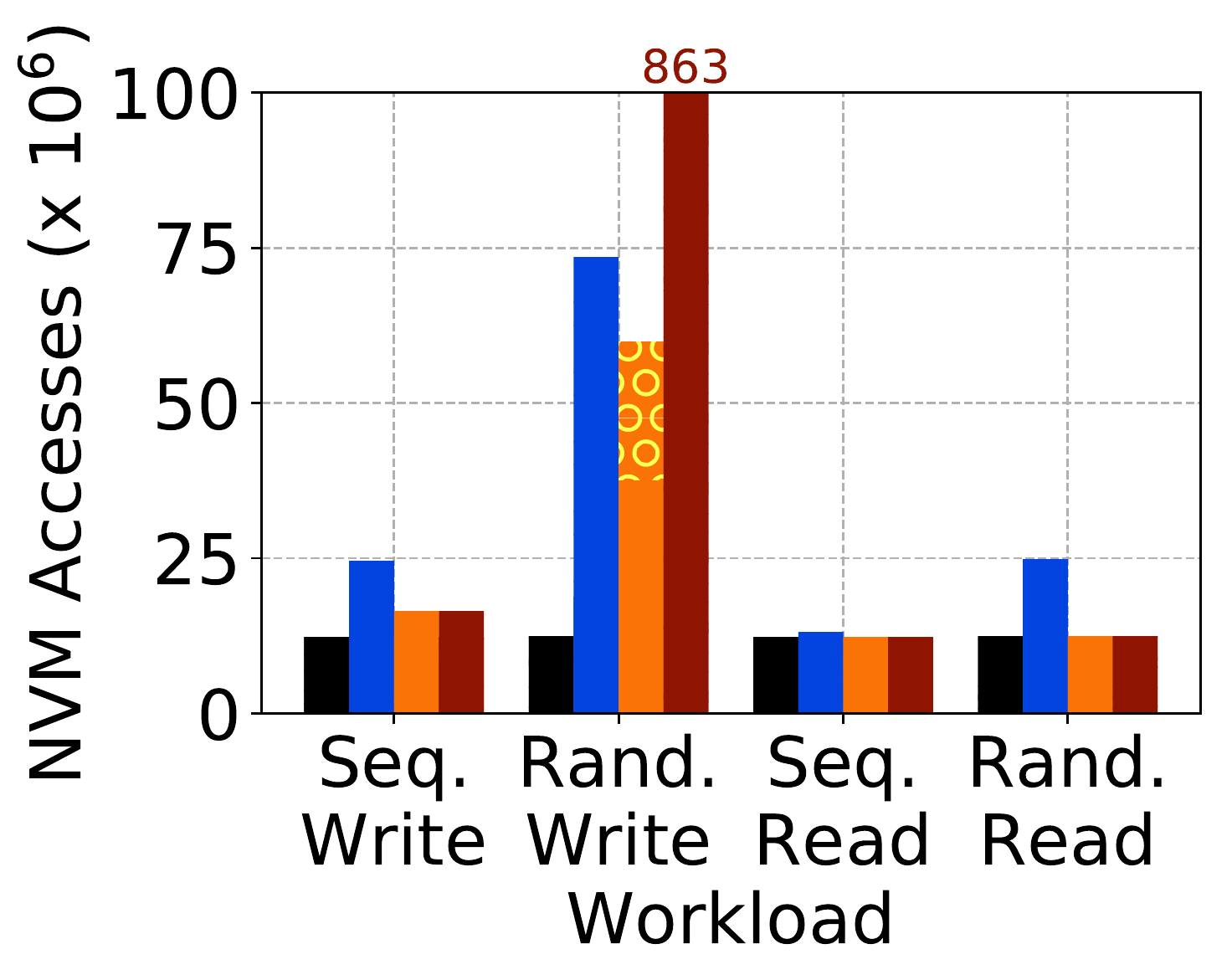}}
		\subfigure[Fio: Cache Accesses]{\label{fig:fio-cache-accesses}
			\includegraphics[width=0.24\textwidth, keepaspectratio]{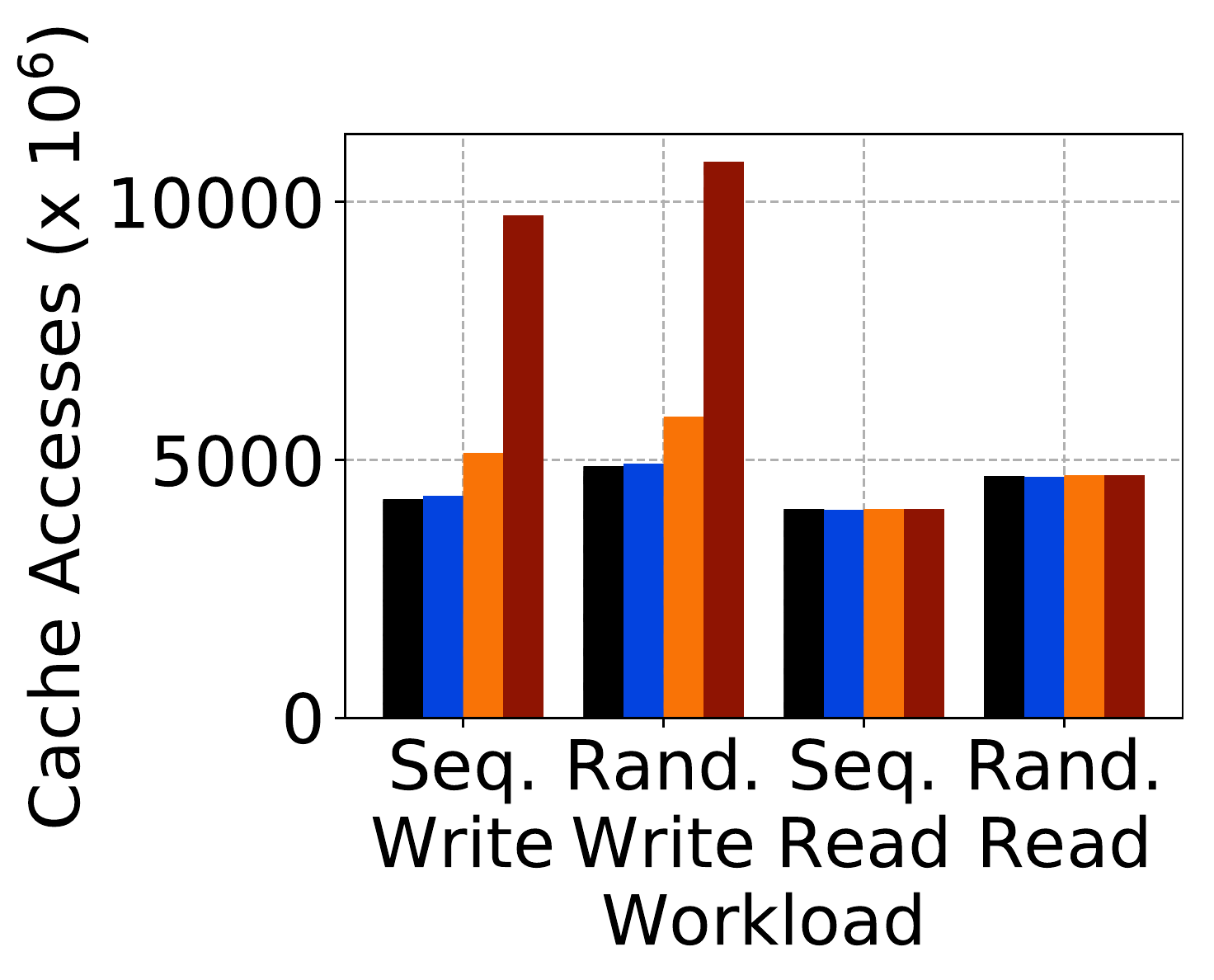}}
		\subfigure[Stream: Runtime]{\label{fig:stream-duration}
			\includegraphics[width=0.24\textwidth, keepaspectratio]{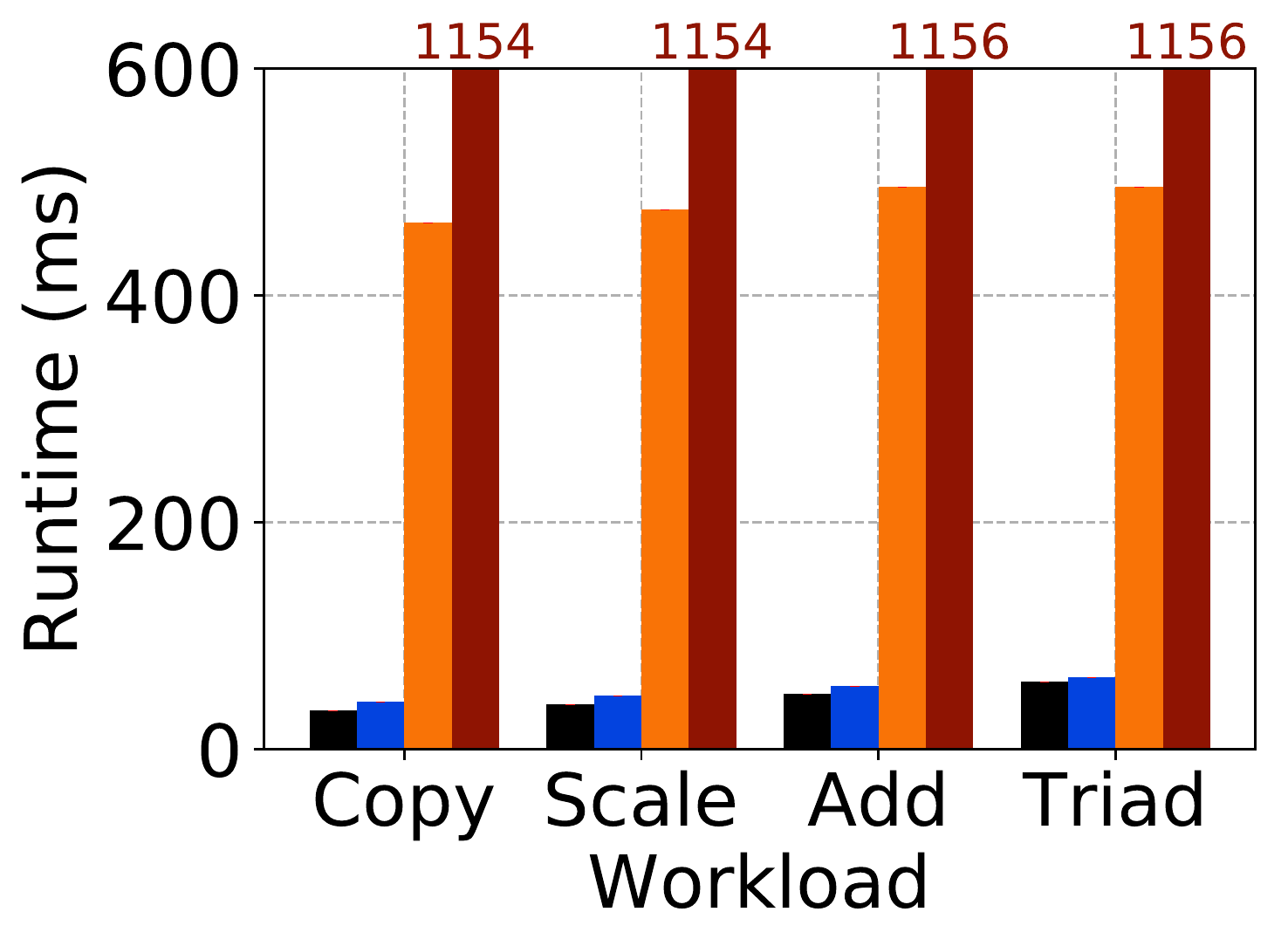}}
		\subfigure[Stream: Energy]{\label{fig:stream-energy}
			\includegraphics[width=0.24\textwidth, keepaspectratio]{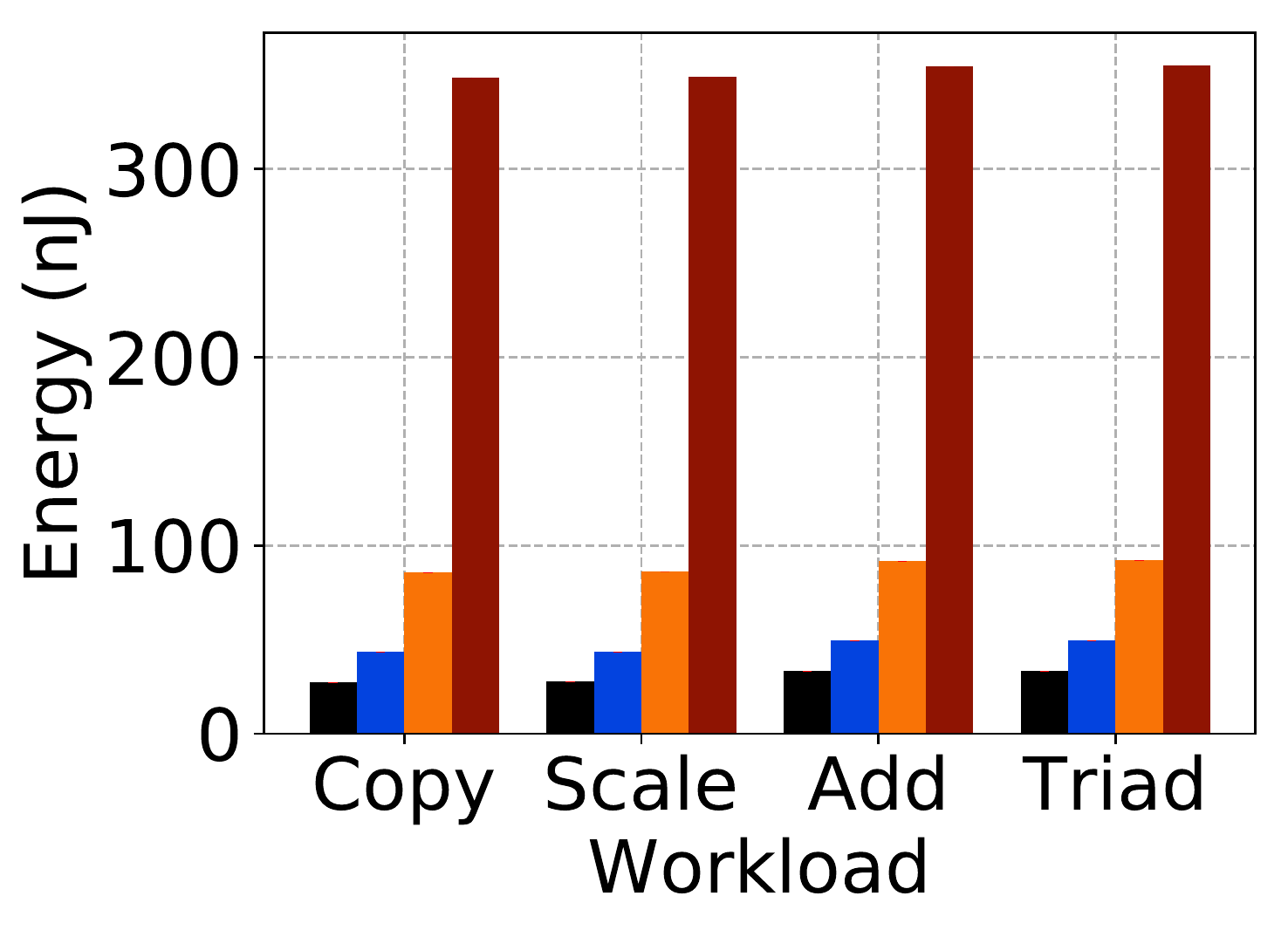}}
		\subfigure[Stream: NVM Accesses]{\label{fig:stream-nvm-accesses}
			\includegraphics[width=0.24\textwidth, keepaspectratio]{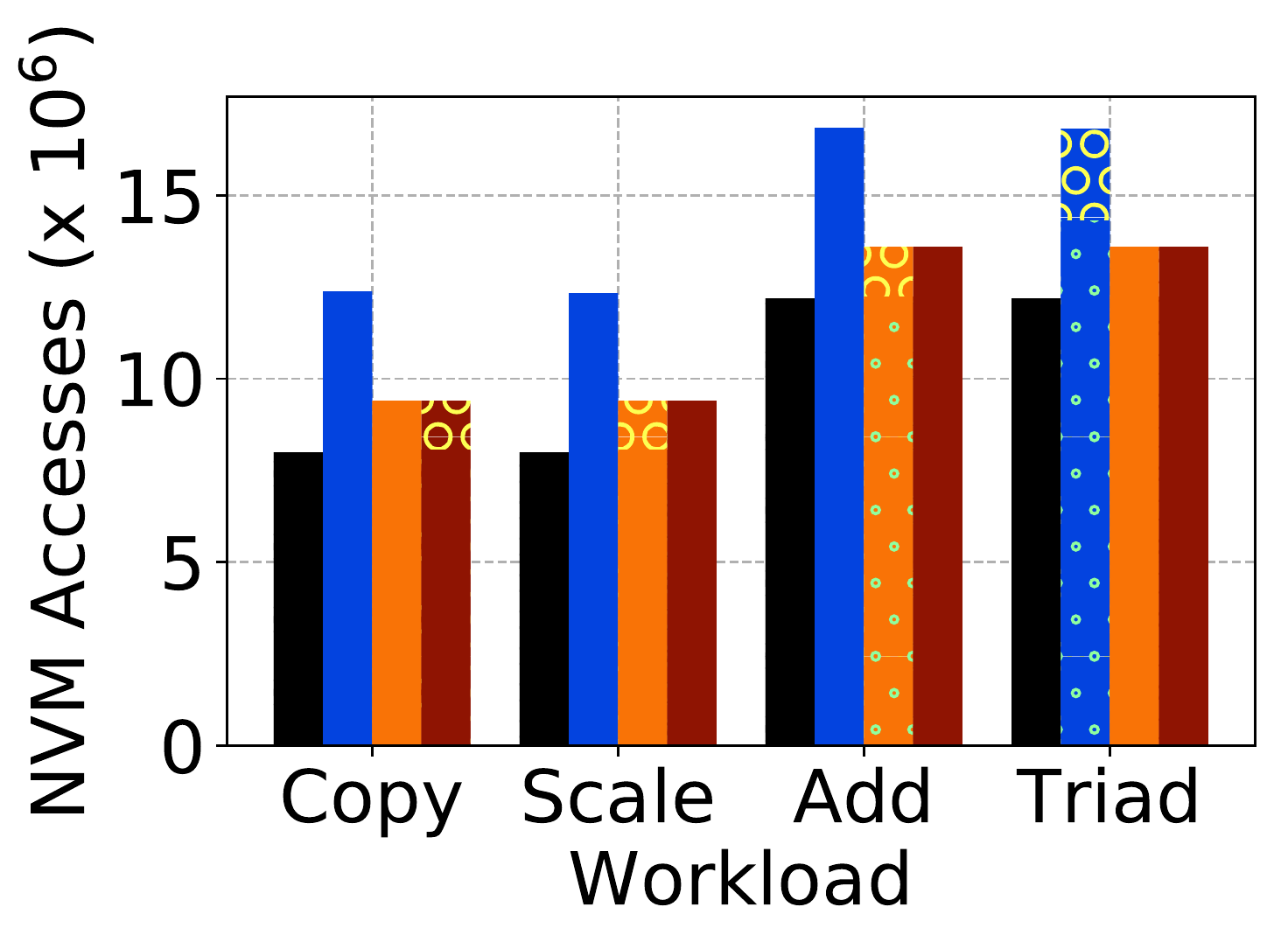}}
		\subfigure[Stream: Cache Accesses]{\label{fig:stream-cache-accesses}
			\includegraphics[width=0.24\textwidth, keepaspectratio]{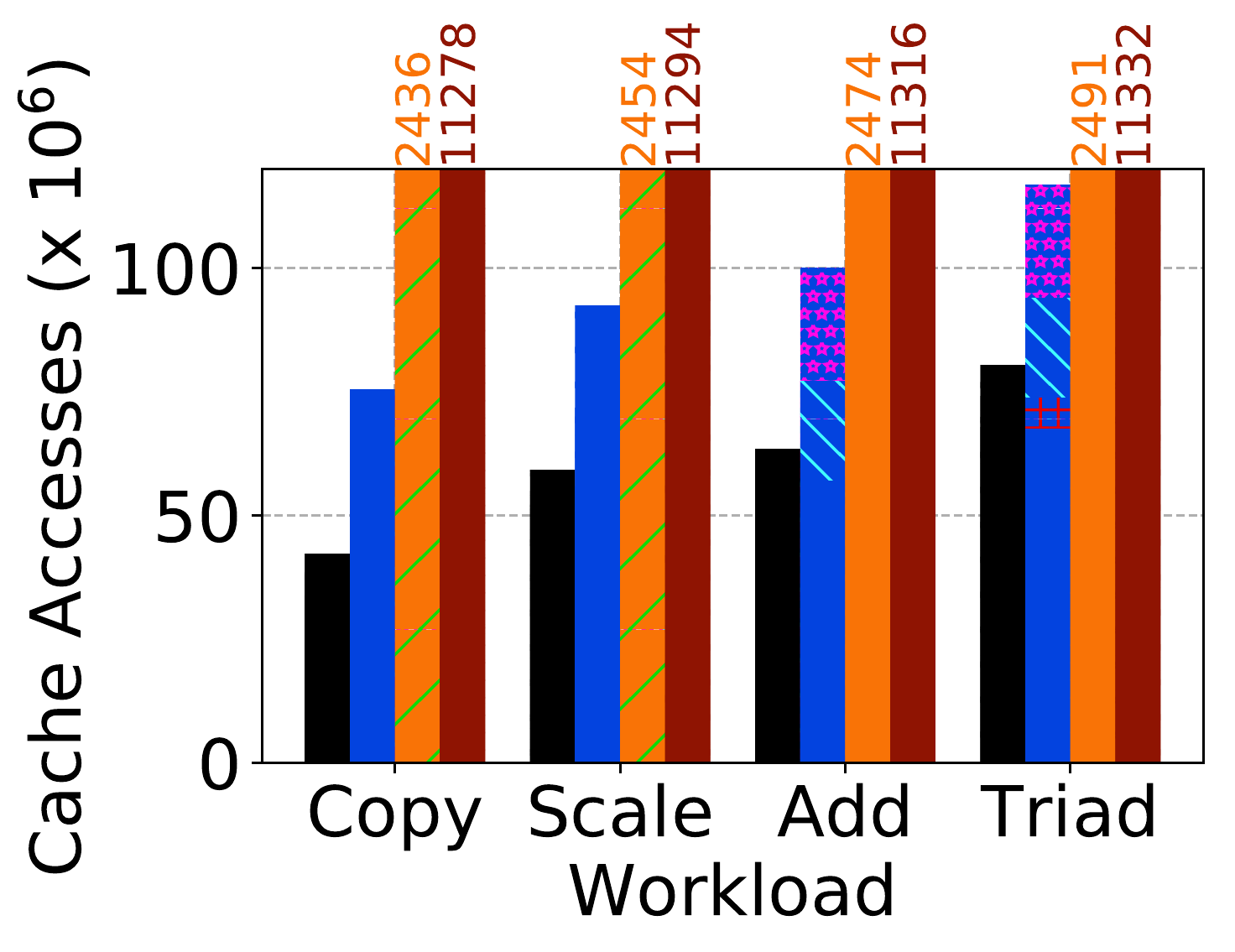}}
	 \end{subfigmatrix}

	 \caption{Runtime, energy, and NVM and cache accesses for various 
		 Redis (\cref{fig:redis-duration,fig:redis-energy,fig:redis-nvm-accesses,fig:redis-cache-accesses}), 
	 tree-based key-value data 
	 structures (\cref{fig:kv-duration,fig:kv-energy,fig:kv-nvm-accesses,fig:kv-cache-accesses}), 
	 N-Store (\cref{fig:nstore-duration,fig:nstore-energy,fig:nstore-nvm-accesses,fig:nstore-cache-accesses}), 
	 Fio (\cref{fig:fio-duration,fig:fio-energy,fig:fio-nvm-accesses,fig:fio-cache-accesses}), 
	 and Stream (\cref{fig:stream-duration,fig:stream-energy,fig:stream-nvm-accesses,fig:stream-cache-accesses})
	 workloads.
	 We divide NVM accesses into data and redundancy information accesses, and 
	 cache accesses into L1, L2, LLC, and on-\system cache.}
\label{fig:results}
\end{figure*}

\newcommand{\redis}{Redis\xspace}
\subsection{Redis}
\label{sec:redis}
Redis is a widely used single-threaded 
in-memory key-value store that uses a 
hashtable as its primary data structure~\cite{redis}. 
We modify Redis (v3.1) to use a persistent memory 
heap using Intel PMDK's libpmemobj 
library~\cite{pmemlib}, building upon an open-source 
implementation~\cite{redis-pmem}. We vary the 
number of Redis instances, each of which operate 
independently. We use the
redis-benchmark utility to spawn 100 clients 
that together generate 1 million requests per Redis instance. 
We use set-only and get-only workloads. 
We show the results only for 6 Redis instances for 
ease or presentation; the trends are the same 
for 1--6 Redis instances that we evaluated. 

\cref{fig:redis-duration} shows the runtime for Redis 
set-only and get-only workloads. 
In comparison to baseline, that maintains no 
redundancy, \system increases the runtime by 
only 3\% for both the workloads. 
In contrast,  
\txbo typically increases the runtime by 50\% 
and \txbp by 200\% over the baseline 
for the set-only workload.
For get-only workloads, 
\txbo and \txbp increase the runtime for 
by a maximum of 
5\% and 28\% over baseline, respectively.
This increase for \txbo and \txbp, 
despite them not verifying any application data reads, 
is because Redis use libpmemobj transactions for 
get requests as well; these transactions 
lead to persistent metadata writes (e.g., to 
set the transaction state as started or committed).
Redis uses transactions for get requests 
because it uses an incremental hashing design 
wherein it rehashes its hashtable incrementally 
upon each request. The incremental rehashing 
can lead to writes for
get requests also. 
We do not change Redis' behavior to eliminate 
these transactions to suit our get-only workload
which wouldn't actually trigger a rehashing. 

\cref{fig:redis-energy,fig:redis-nvm-accesses,fig:redis-cache-accesses} 
show the energy, NVM accesses and cache accesses. 
The energy results are similar to that for runtime. 
For the set-only workload, 
\system performs more NVM accesses than \txbo 
because \system does not cache the data or redundancy 
information in the L1 and L2 caches; \txbo instead
performs more cache accesses. Even though \txbp 
can and does use the caches (demonstrated by 
\txbp's more than 200$\times$ more cache accesses than 
baseline), it also requires more NVM accesses because 
it needs to read the entire page to 
compute the page-granular system-checksums.
For get-only workloads, \system performs
more NVM accesses than both \txbo and \txbp
because it verifies the application data reads with \tcsums.

\subsection{Key-value Data Structures}
\label{sec:kv}
We use three persistent memory 
key-value data structures, 
namely C-Tree, B-Tree, and 
RB-Tree, from Intel PMDK~\cite{pmemlib}. 
We use PMDK's pmembench utility to 
generate insert-only, update-only, 
balanced (50:50 updates:reads), and read-only workloads. 
We stress the NVM usage by using 
12 instances of each data-structure; 
each instance is driven by a single 
threaded workload generator. Having 
12 independent instances of 
single-threaded workloads allows us 
to remove locks from the data-structures 
and increase the workload throughput. 
We show the results for insert-only 
and balanced workloads; the trends are the 
same for other workloads.

\cref{fig:kv-duration,fig:kv-energy,fig:kv-nvm-accesses,fig:kv-cache-accesses} 
show the runtime, energy, and NVM and cache accesses for 
the different workloads and data-structures. 
For the insert-only workload, \system 
increases the runtime by a maximum of 1.5\% (for RB-Tree) 
over the baseline while updating the redundancy for all 
inserted tuples. In contrast, 
\txbo and \txbp increase the runtime by 43\% and 171\% over the 
baseline, respectively. 
For the balanced workload, 
\system updates the redundancy for tuple updates and 
also verifies tuple reads with system-checksums with 
only 5\% increase in runtime over the baseline 
for C-Tree and B-Tree. \txbo incurs a 
20\% increase in runtime over baseline for just 
updating the redundancy upon tuple updates; \txbp 
performs even worse. 

\newcommand{\nstore}{N-Store\xspace}
\subsection{N-Store}
\label{sec:n-store}
N-Store is a NVM-optimized relational DBMS. 
We use update-heavy (90:10 updates:reads), 
balanced (50:50 updates:reads) and read-heavy (10:90 updates:reads) 
YCSB workloads with high skew (90\% of transactions go 
to 10\% of tuples)~\cite{nstore-sigmod}. 
We use 4 client threads to drive the workload 
and perform a total of 800000 transactions. 
For N-Store, we present results from a 
single run with no error bars.

\cref{fig:nstore-duration,fig:nstore-energy,fig:nstore-nvm-accesses,fig:nstore-cache-accesses} 
show runtime and energy, and NVM and cache accesses. 
\system increases the runtime by 27\% and 
41\% over the baseline for the 
read-heavy and update-heavy workloads, respectively. 
\system's overheads are higher with N-Store,
than with Redis or key-value structures, because N-Store 
uses a linked list based write-ahead log that leads to a 
random write access pattern for update transactions. 
Each update transaction allocates and writes to a
linked list node. Because the linked list layout 
is not sequential in NVM, \system incurs cache-misses 
for the redundancy information and performs
more NVM accesses. The random write 
access pattern also affects \txbo and \txbp, 
with a 70\%--117\% and 
264\%--600\% longer runtime than baseline, 
respectively. 
This is because the \txbo and \txbp also 
incur misses for for redundancy information 
in the L1, L2 and LLC caches and have to 
perform more NVM accesses for random writes. 

\newcommand{\fio}{Fio\xspace}
\subsection{Fio Benchmarks}
\label{sec:fio}
Fio is a file system benchmarking tool that supports
multiple access patterns~\cite{fio}. 
We use \fio's libpmem engine that accesses 
DAX-mapped NVM file data using load and 
store instructions. We use sequential and random 
read and write workloads with a 64B access granularity. 
We use 12 concurrent threads with each thread performing 
32MB worth of accesses (reads or writes). Each thread 
accesses data from a non-overlapping 512MB region, and 
no cache-line is accessed twice. 

\cref{fig:fio-duration,fig:fio-energy,fig:fio-nvm-accesses,fig:fio-cache-accesses} 
show the results for fio. 
As discussed above in the context of N-Store, 
random access pattern in the application hurt \system 
because of poor reuse for redundancy cache-lines with 
random accesses. 
This trend is visible for fio as well---whereas 
\system has essentially the same runtime as 
baseline for sequential accesses, \system increases 
the runtime by 2\% and 33\% over baseline 
for random reads and writes, respectively. 
However, \system still outperforms \txbo and 
\txbp for the write workloads. For read workloads, 
\txbo and \txbp have no impact because they 
do not verify application data reads. 
For the random write 
workload, \system incurs a higher energy overhead 
than \txbo. This is because the energy required for 
additional NVM accesses that \system generates 
exceed that required for the additional 
cache accesses that \txbo generates. 

\subsection{Stream Benchmarks}
\label{sec:stream}
Stream is a memory bandwidth stress tool~\cite{stream} 
that is part of the HPC Challenge suite~\cite{hpcc-bench}. 
Stream comprises of four sequential access kernels: 
(i) \emph{Copy} data from one array to another,
(ii) \emph{Scale} elements from one array by a 
constant factor and write them in a second array, 
(iii) \emph{Add} elements from two arrays and write them 
in a third array, and 
(iv) \emph{Triad} which is a combination of 
Add and Scale: it scales the elements from one array, adds 
them to the corresponding elements from the second array,
and stores the values in a third array.
We modify stream to store and access data in persistent 
memory. We use 12 concurrent threads that 
operate on non-overlapping regions of the 
arrays. Each array has a size of 128MB.

\cref{fig:stream-duration,fig:stream-energy,fig:stream-nvm-accesses,fig:stream-cache-accesses} 
show the results for the four kernels. The trends are similar to 
the preceding results. \system, \txbo, and \txbp 
increase the runtime by 
6\%--21\%, 700\%--1200\%, and 1800\%--3200\% over the 
baseline, respectively. 
The absolute value of the overheads are higher for all the 
designs because the baseline already saturates the NVM bandwidth, 
unlike the real-world applications considered above that consume 
the data in more complex fashions. 
The impact of computation complexity is clear even across
the four microbenchmarks: copy is the simplest kernel, 
followed by scale, add, and triad. Consequently, 
the overheads for all the designs are highest for 
the copy kernel and lowest for the triad kernel.

\subsection{Impact of \system's Design Choices}
\label{sec:benefit-breakdown}
We break down the impact of \system's design choices, 
namely, using \tcsum, caching redundancy information, 
and storing data diff in LLC. 
We present the 
results for one 
workload from each of the above applications: 
set-only workload with 6 instances for Redis, 
insert-only workload for C-Tree, balanced workload for N-Store, 
random write workload for fio, and triad 
kernel for stream.

\begin{figure}[t]
	\centering
	\framebox[0.7\columnwidth][c]{
	\includegraphics[width=0.7\columnwidth, keepaspectratio]{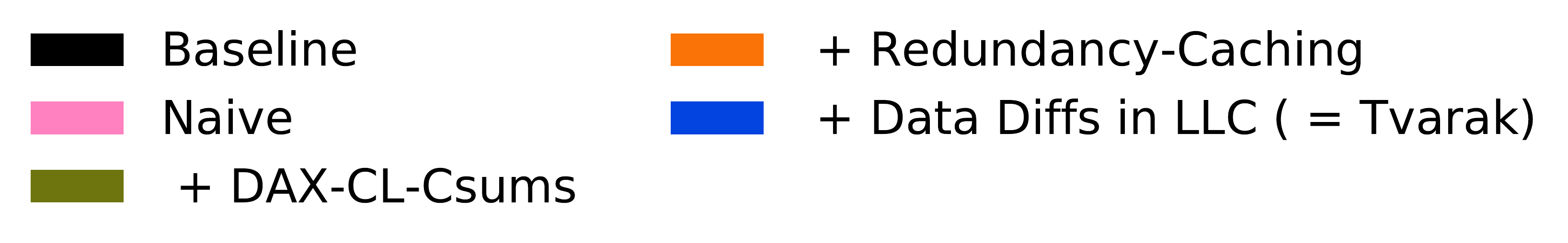}
	}
	{\includegraphics[width=0.7\columnwidth]{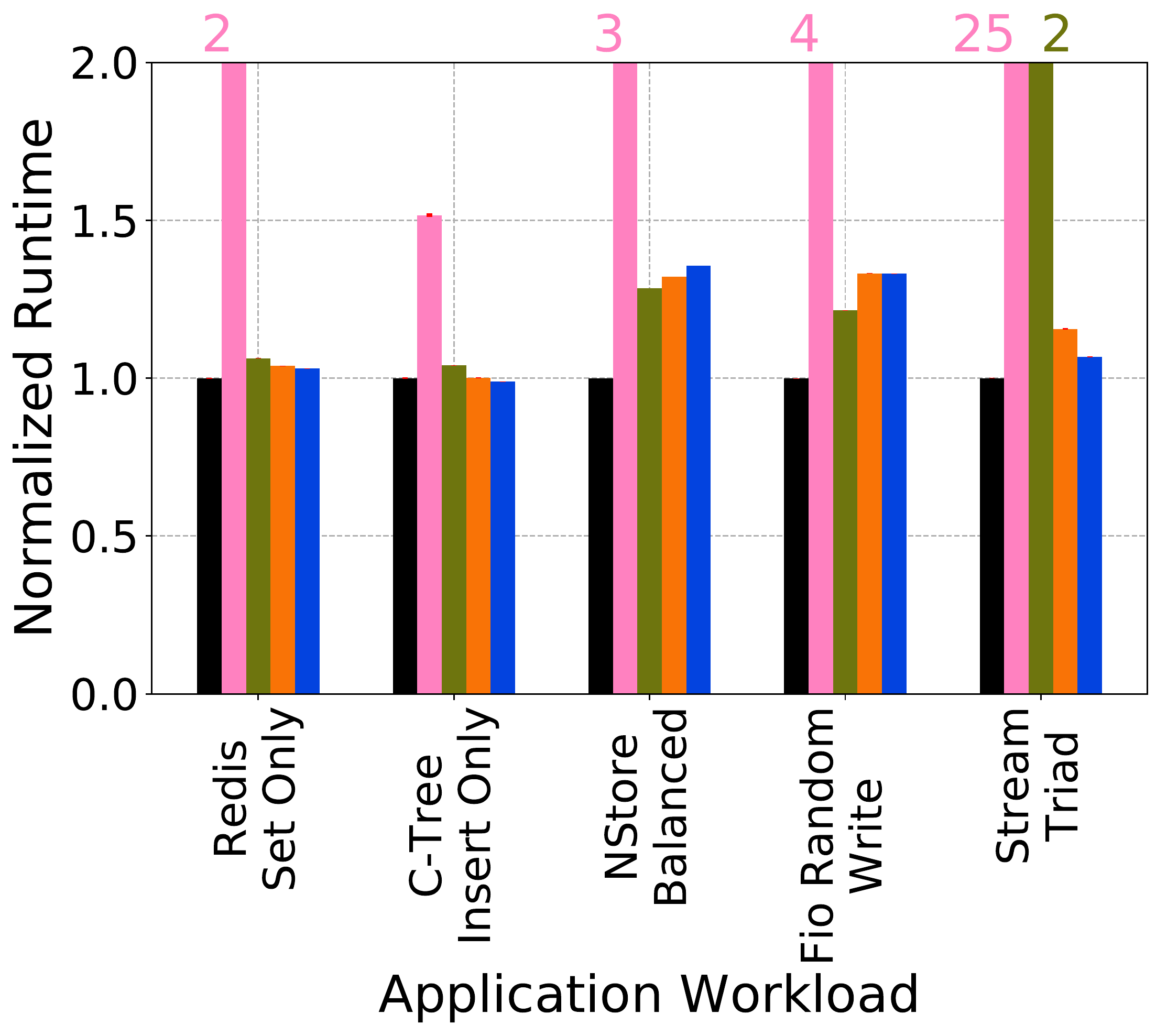}}
	\caption{\textbf{Impact of \system's Design Choices}:
		We evaluate the impact of \system's design optimizations 
		with one workload for each application. We present the 
		results for the naive design and then add 
		optimizations, \tcsums, 
		redundancy caching, 
		and data diffs in LLC. With all the 
		optimizations enabled, we get \system.}
	\label{fig:benefit-breakdown}
\end{figure}

\cref{fig:benefit-breakdown} shows the performance 
for the naive design, and then adds individual 
design elements, i.e., \tcsums, redundancy caching, 
and storing data diffs in LLC. With all the 
design elements, we get the complete \system design. 
For Redis, C-Tree and stream's triad kernel, all 
of \system's design choices improve performance.
This is the case for B-Tree, RB-Tree, other stream 
kernels, and fio sequential access workloads as 
well (not shown in the figure).
For N-Store and fio random write workload, 
redundancy caching and storing data diffs in the 
LLC hurt performance. This is because 
taking away cache space from application data 
creates more NVM accesses than that saved by caching the 
redundancy data and storing the data diffs in LLC for N-Store 
and fio random writes:w
---their 
random access patterns lead to poor reuse of 
redundancy cache-lines.

This evaluation highlights the importance of 
choosing the LLC partition space that 
\system uses to cache 
redundancy information or to store data diffs. 
We leave dynamically adapting the partition sizes 
based on the workload characteristics for 
future work. The partition sizes can be 
adapted either by \system using 
set duelling~\cite{set-duelling-isca}, 
or by the OS by application profiling. 

\subsection{Sensitivity Analysis}
\label{sec:sensitivity}
We evaluate the sensitivity of \system to the 
size of LLC partitions that it can use for 
caching redundancy information and storing data diffs. 
We present the results for one workload 
from each of the set of applications, namely, 
set-only workload with 6 instances for Redis, 
insert-only workload for C-Tree, balanced workload for N-Store, 
random write workload for fio, and triad 
kernel for stream.

\begin{figure}[t]
		\centering
	\framebox[0.7\columnwidth][c]{
	\includegraphics[width=0.7\columnwidth, keepaspectratio]{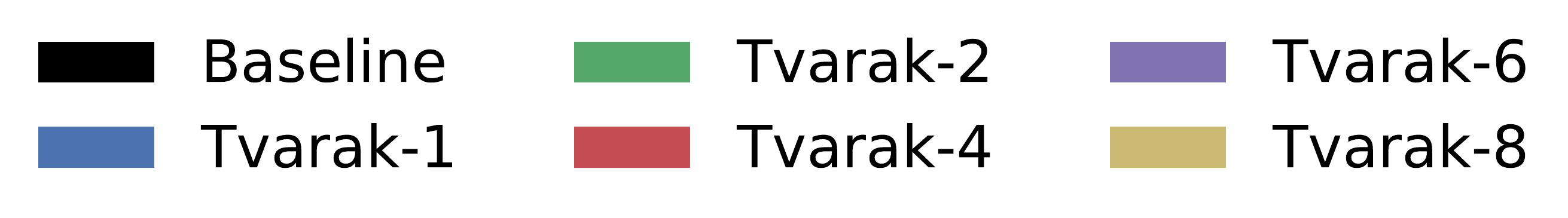}
	}
	\\
	 \begin{subfigmatrix}{2}
		 \subfigure[Sensitivity to number of ways for caching redundancy information.]
			{\label{fig:redundancy-sensitivity}
			\includegraphics[height=1.5\linewidth, keepaspectratio]{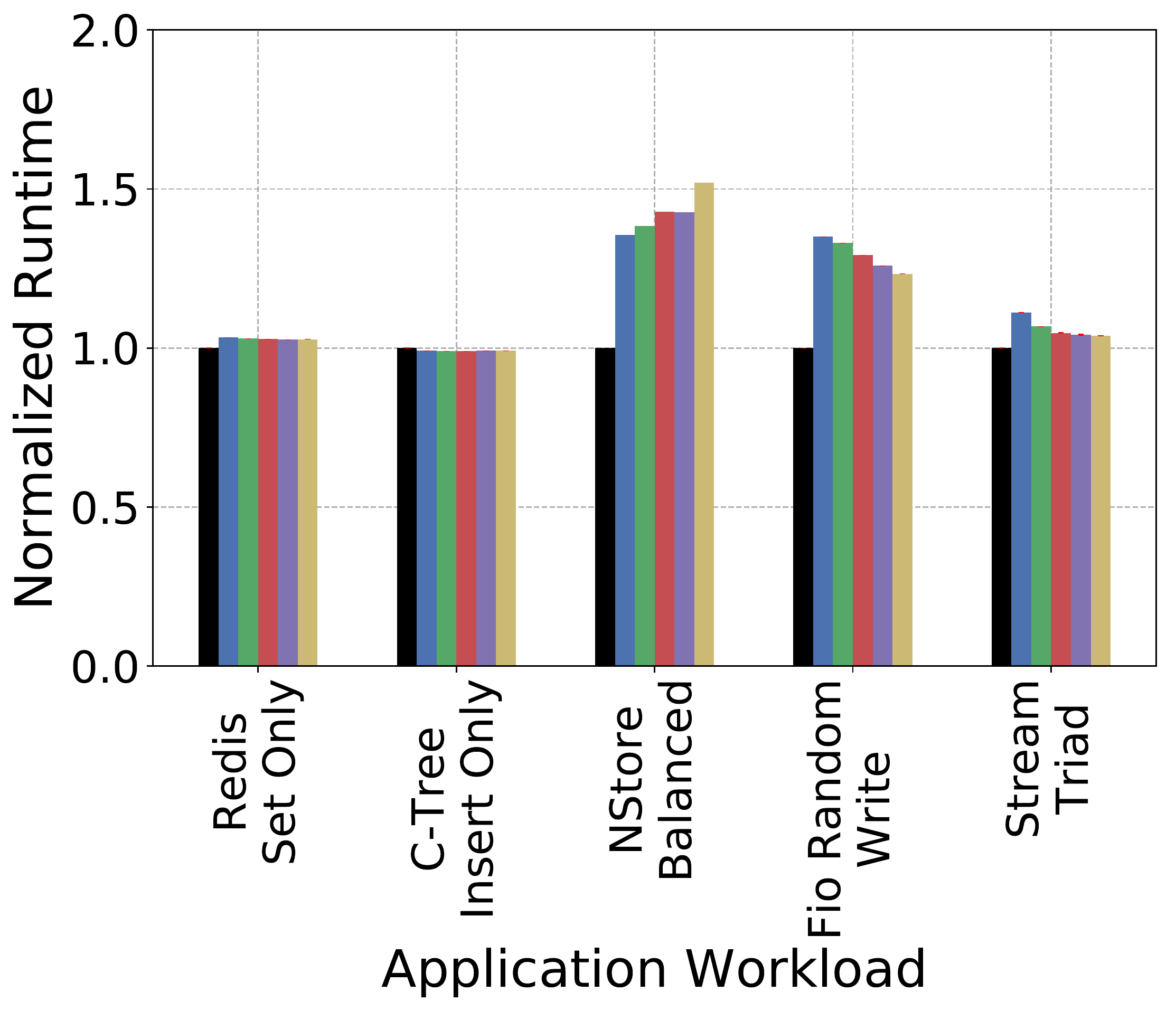}}
		 \subfigure[Sensitivity to number of ways for storing data diffs.]
			{\label{fig:diffdata-sensitivity}
			\includegraphics[height=1.5\linewidth, keepaspectratio]{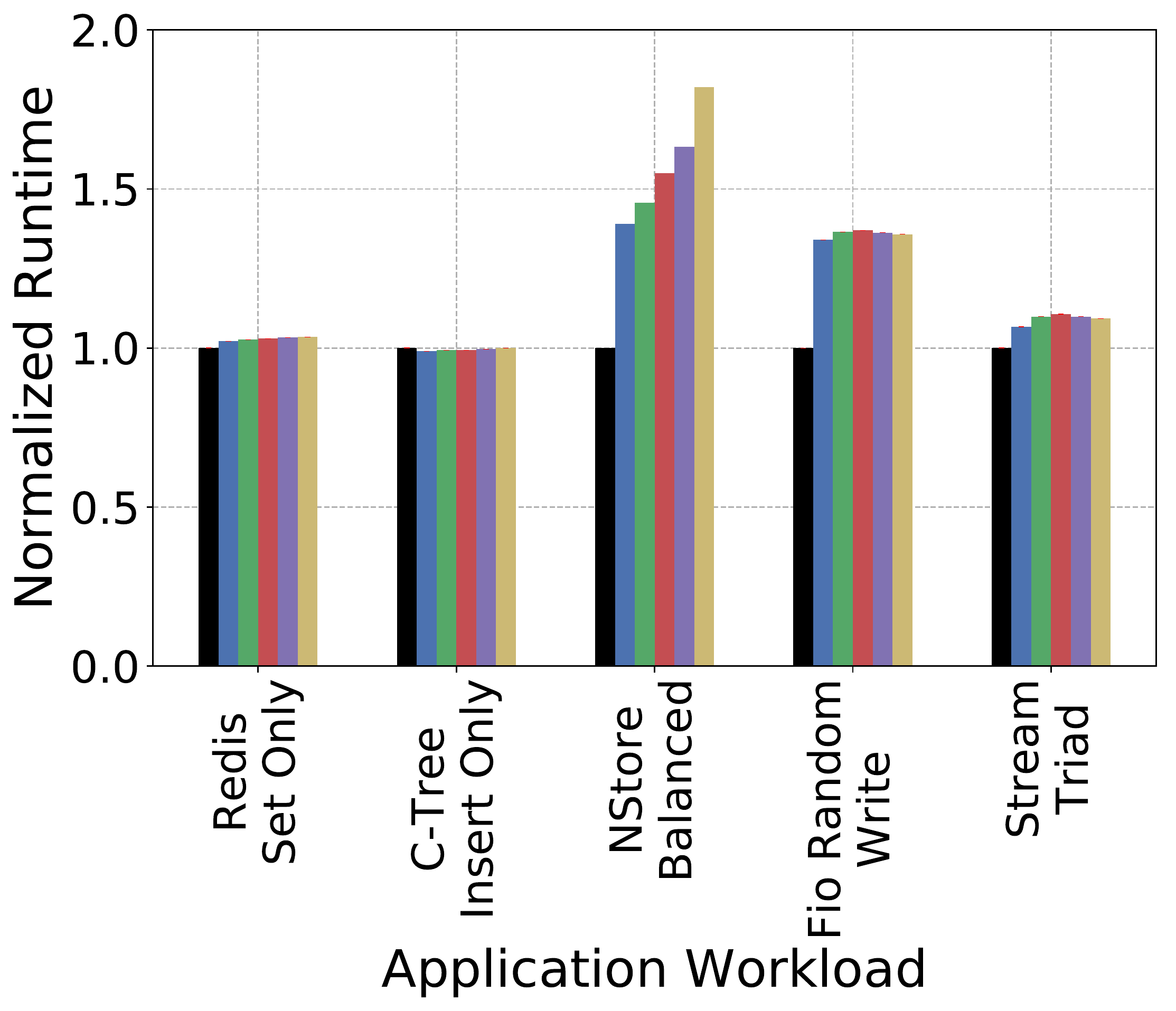}}
	 \end{subfigmatrix}
	 \caption{Impact of changing the number of LLC ways (out of 16) 
		 that \system can use for caching redundancy data and for storing data diffs.}
\label{fig:llc-sensitivity}
\end{figure}

\cref{fig:redundancy-sensitivity} shows the 
impact of changing the number of LLC ways 
(out of 16) that \system can use for caching 
redundancy information. 
Redis and C-Tree are largely unaffected by the 
redundancy partition size, with 
Redis benefitting marginally from 
reserving 2 ways instead of 1. 
Stream and fio, being synthetic 
memory stressing microbenchmarks, 
demonstrate that dedicating a larger 
partition for redundancy caching improves \system's 
performance because of the increased 
cache space. N-Store is cache-sensitive 
and taking away the cache from
application data for redundancy 
hurts its performance. 

\cref{fig:diffdata-sensitivity} shows the sensitivity 
of \system to the number of ways reserved for 
storing data diffs. As with the sensitivity 
to redundancy information partition size, 
changing the data diff partition size has 
negligible impact on Redis and C-Tree. 
For N-Store, increasing the number of 
ways reserved for storing data diffs 
hurts performance because N-Store is 
cache-sensitive. Stream and fio show an interesting pattern, 
increasing the number of data diff ways from 
1 to 4 hurts performance, but increasing it 
to 6 or 8 improves performance (although 
the performance remains worse than 
reserving just 1 way).
This is because dedicating more 
ways for storing data diffs has two 
contradicting effect. It reduces the number 
of write-backs due to data diff evictions, 
but it also causes more write-backs because of 
the reduced cache space for application data. 
Their combined effect dictates the overall performance. 

We also evaluate the impact of increasing the 
number of NVM DIMMs and changing the underlying 
NVM technology on baseline, \system, \txbo, and \txbp.
The relative performance trends stay the same 
with both of these changes; we do not show the 
results here for brevity. As an example, 
even with 8 NVM DIMMs or 
with improved NVM performance by 
considering battery-backed DRAM as NVM, 
\system continues to outperform
\txbo and \txbp by orders of magnitude for 
the stream microbenchmarks. 

%% file: tables/workloads.tex
	\scriptsize {
\begin{tabular}{r l}
	\makecell[r]{Redis} & 
	\makecell[l]{Set-only and get-only with 1--6 parallel instances}\\
\hline
	\makecell[r]{C-Tree} & 
	\makecell[l]{Insert-only, and 100:0, 50:50, \& 0:100 updates:reads with 12 parallel instances}\\
\hline
	\makecell[r]{B-Tree} & 
	\makecell[l]{Insert-only, and 100:0, 50:50, \& 0:100 updates:reads with 12 parallel instances}\\
\hline
	\makecell[r]{RB-Tree} & 
	\makecell[l]{Insert-only, and 100:0, 50:50, \& 0:100 updates:reads with 12 parallel instances}\\
\hline
	\makecell[r]{Fio} & 
	\makecell[l]{Sequential and random reads and writes with 12 threads}\\
\hline
	\makecell[r]{Stream} & 
	\makecell[l]{4 memory bandwidth intensive kernels with 12 threads}\\
\end{tabular}
	}

%% file: tables/sim-params.tex
	\scriptsize {
\begin{tabular}{r l}
	Cores & 
	\makecell[l]{12 cores, x86-64 ISA, 2.27 GHz,\\Westmere-like OOO~\cite{zsim-isca}}\\[1.5ex]
	\hline\\[-1ex]
	L1-D caches &
	\makecell[l]{32KB, 8-way set-associative, 4 cycle latency,\\
	LRU replacement, 15/33 pJ per hit/miss~\cite{cacti-micro}}\\\\
	L1-I caches &
	\makecell[l]{32KB, 4-way set-associative, 3 cycle latency,\\
	LRU replacement, 15/33 pJ per hit/miss~\cite{cacti-micro}}\\\\
	L2 caches &
	\makecell[l]{256KB, 8-way set-associative, 7 cycle latency,\\
		LRU replacement, 46/94 pJ per hit/miss~\cite{cacti-micro}}\\\\
	L3 cache &
	\makecell[l]{24MB (12 2MB banks), 16-way set-associative,\\ 
	27 cycle latency, shared and inclusive,\\MESI coherence, 64B lines\\
	LRU replacement, 240/500 pJ per hit/miss~\cite{cacti-micro}}\\[1.5ex]
\hline\\[-1ex]
	DRAM & 6 DDR DIMMs, 15ns reads/writes \\\\
	NVM & \makecell[l]{4 DDR DIMMs, 60ns reads, 150ns writes~\cite{pcm-isca}\\
	1.6/9 nJ per read/write~\cite{pcm-isca}}\\[1.5ex]
\hline\\[-1ex]
	\system &
	\makecell[l]{4KB on-controller cache with 1 cycle latency, 15/33 pJ per hit/miss\\
		2 cycle latency for address range matching\\
		1 cycle per checksum/parity computation and verification,\\
		2 ways (out of 16) reserved for caching redundancy information,\\
		1 way (out of 16) for storing data diffs. 
}\\
\end{tabular}
	}

%% file: conclusion.tex
\section{Conclusion}
\label{sec:conclusion}

\system efficiently maintains system-checksums and cross-device parity
for DAX NVM storage, addressing controller and firmware imperfections
expected to arise with NVM as they have with other storage technologies.
As a hardware offload, managed by the storage software, \system does so with
minimal overhead and much more efficiently that software-only approaches.
Since system-level redundancy is expected from production storage,
\system{} is an important step towards the use of DAX NVM as primary 
storage.